\documentclass[preprint,eqsecnum,aps,nofootinbib]{revtex4}
\usepackage{amsfonts,amsmath,amssymb,amsthm}
\usepackage{latexsym}
\usepackage{bbm,bm}
\usepackage{graphicx}

\newcommand{\ket}[1]{\lvert #1 \rangle}
\newcommand{\bra}[1]{\langle #1 \lvert}
\newcommand{\beq}{\begin{equation}}
\newcommand{\eeq}{\end{equation}}
\newcommand{\beqs}{\begin{eqnarray}}
\newcommand{\eeqs}{\end{eqnarray}}

\begin{document}

\title{Greenberger-Horne-Zeilinger Symmetry in Four Qubit System}

\author{DaeKil Park$^{1,2}$}

\affiliation{
             $^1$Department of Electronic Engineering, Kyungnam University, Changwon
                 631-701, Korea       \\
             $^2$Department of Physics, Kyungnam University, Changwon
                  631-701, Korea                                            
             }

\begin{abstract}
Like a three-qubit Greenberger-Horne-Zeilinger(GHZ) symmetry we explore a corresponding symmetry in the four-qubit system, which 
we call GHZ$_4$ symmetry. While whole GHZ-symmetric states can be represented by two real parameters,  the whole set of the 
GHZ$_4$-symmetric states is represented by three real parameters. In the parameter space 
all GHZ$_4$-symmetric states reside inside a tetrahedron. 
We also explore a question where the given SLOCC class of the GHZ$_4$-symmetric states resides in the tetrahedron.
Among nine SLOCC classes we have examined five SLOCC classes, which results in three linear hierarchies 
$L_{abc_2} \subset L_{a_4} \subset L_{a_2b_2} \subset G_{abcd}$, $L_{a_20_{3\oplus\bar{1}}} \subset G_{abcd}$, and
$L_{0_{3\oplus\bar{1}}0_{3\oplus\bar{1}}} \subset G_{abcd}$ which hold, at least, in the whole set of the GHZ$_4$-symmetric states.
Difficulties arising in the analysis of the remaining SLOCC classes are briefly discussed.
\end{abstract}

\maketitle

\section{Introduction}
Quantum entanglement\cite{text,horodecki09} is the most important notion in quantum mechanics and quantum information theory.
Research into quantum entanglement was initiated from the very beginning of quantum mechanics\cite{epr-35,schrodinger-35}. At that time the main motivation for the study was pure theoretical in the context of the nonlocal properties of quantum mechanics. However, recent considerable attention to it is 
mainly due to its crucial role as a physical resource in various quantum information processing.
In fact, quantum entanglement plays a central role in quantum teleportation\cite{teleportation},
superdense coding\cite{superdense}, quantum cloning\cite{clon}, and quantum cryptography\cite{cryptography,cryptography2}. It is also quantum entanglement, which makes the quantum computer\footnote{The current status of quantum computer technology was reviewed in Ref.\cite{qcreview}.} outperform the classical one\cite{computer}. Thus, it is essential to understand how to quantify 
and how to characterize the multipartite entanglement. Still, however, this issue is not completely understood.

The most direct classification of the multipartite entanglement is to use the local unitary (LU), i.e., the unitary operations acted independently on each of the subsystems.
Since quantum entanglement is a nonlocal property of a given multipartite state, it should be invariant under the LU transformations.
The LU transformation is related to local operations and classical communication 
(LOCC) \cite{bennet00-1,vidal00} as follows. Let two quantum states, say $\ket{\psi}$ and $\ket{\varphi}$, be in the same category of LU. 
Then, one state can be converted into the other one with certainty by means of LOCC. Although the LU is a useful tool for the classification of the multipartite 
entanglement, it generates infinite equivalent classes even in the simplest bipartite systems.

In order to escape this difficulty the authors in Ref. \cite{bennet00-1} suggested the classification through stochastic local operations and classical 
communication (SLOCC). If $\ket{\psi}$ and $\ket{\varphi}$ are in the same SLOCC class, this means that one state can be converted into the 
other state with nonzero probability by means of LOCC. Mathematically, if two $n$-party states $\ket{\psi}$ and $\ket{\varphi}$ are in the same SLOCC class, they are related to each other by 
$\ket{\psi} = A_1 \otimes A_2 \otimes \cdots \otimes A_n \ket{\varphi}$ with $\{A_j\}$ being arbitrary invertible local 
operators\footnote{For complete proof on the connection between SLOCC and local operations see Appendix A of Ref.\cite{dur00}.}. 
However, it is more useful to restrict ourselves to the SLOCC transformation where all $\{A_j\}$ belong to 
SL($2$, $C$), the group of $2 \times 2$ complex matrices having determinant equal to $1$. 

The SLOCC classification was first examined in the three-qubit pure-state system\cite{dur00}. It was shown that the whole system consists of 
six inequivalent SLOCC classes, i.e., fully separable (S), three bi-separable (B), W, and Greenberger-Horne-Zeilinger (GHZ) classes.
Moreover, it is possible to know which class an arbitrary state $\ket{\psi}$ belongs 
by computing the residual entanglement $\tau_3 (\psi)$\cite{ckw} and concurrences ${\cal C} (\psi)$\cite{concurrence1} for its partially reduced states. Similarly, 
the entanglement of whole three-qubit mixed states also consists of S, B, W, and GHZ types\cite{threeM}. It was shown that these classes 
satisfy a linear hierarchy S $\subset$ B $\subset$ W $\subset$ GHZ. 

Although SLOCC classes for the three-qubit system are well-known, still it is highly difficult problem to know which type of entanglement 
is contained for arbitrary three-qubit mixed states. This is mainly due to the fact that the analytic computation of the residual entanglement 
for arbitrary mixed state is generally impossible except few rare case\cite{tangle}. Recently, a significant progress has been made in 
this issue in Ref. \cite{elts12-1}.  Authors in Ref. \cite{elts12-1} examined the whole set of the three-qubit GHZ-symmetric states. This is an 
invariant symmetry under (i) qubit permutations, (ii) simultaneous flips, (iii) qubit rotations about the $z$-axis. 
It was shown that the whole GHZ-symmetric states can be parametrized by two real parameters, {\it say} $x$ and $y$.
\begin{figure}[ht!]
\begin{center}
\includegraphics[height=8cm]{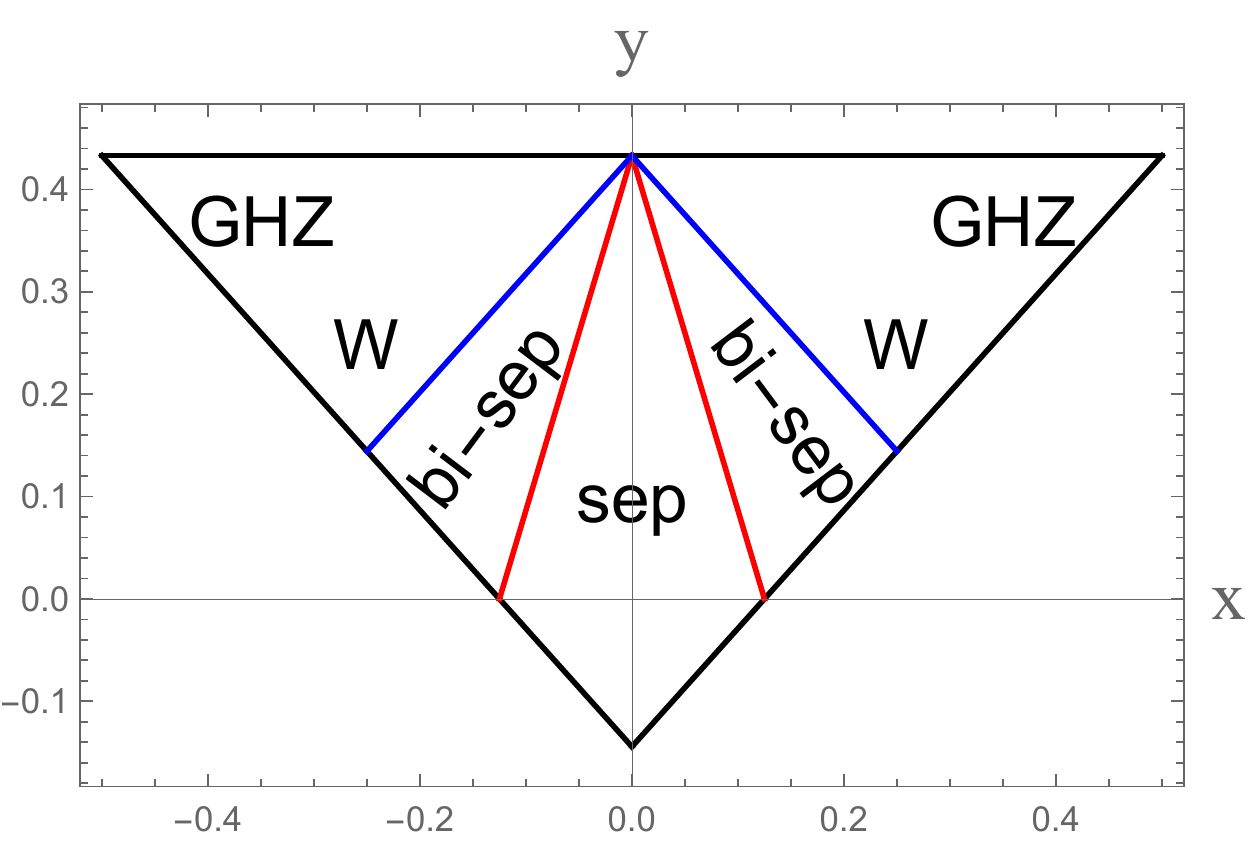}
\caption[fig1]{(Color online) Complete classification of GHZ-symmetric states.}
\end{center}
\end{figure}
The whole GHZ-symmetric states are represented as points inside a triangle of Fig. 1 in $x-y$ plane.
Authors in Ref. \cite{elts12-1}
succeeded in classifying the entanglement of the three-qubit GHZ-symmetric states completely. The result is shown in Fig. 1, where the
 linear hierarchy S $\subset$ B $\subset$ W $\subset$ GHZ holds in this subset states.  
 This complete classification makes it possible to 
compute the three-tangle, square root of the residual entanglement, analytically 
for the whole GHZ-symmetric states\cite{siewert12-1} and to construct the class-specific optimal witnesses\cite{elts12-2}.
It also makes it possible to obtain lower bound of three-tangle for arbitrary three-qubit mixed state\cite{elts13-1}.  
More recently, the SLOCC classification of the extended GHZ-symmetric states was discussed\cite{jung13-1}. Extended GHZ symmetry is the GHZ symmetry without qubit permutation symmetry. Thus, it is larger symmetry group than usual GHZ symmetry group, and is parametrized by four real parameters.

\begin{center}
\begin{tabular}{c|cccccc} \hline \hline
 References &  Ref. \cite{dur00} &  Ref. \cite{fourP-1} &  Ref. \cite{fourP-2} &  Ref. \cite{fourP-3} &  Ref. \cite{fourP-4} &  Ref. \cite{fourP-5}   \\  \hline 
 number of SLOCC classes & $\infty$ & $9$ & $8$ & $23$ & $49$ & $9$              \\   \hline  \hline
\end{tabular}

\vspace{0.2cm}
Table I: Number of SLOCC classes of four-qubit pure states in various references.
\end{center}

The SLOCC classification of the four-qubit system was explored in Ref. \cite{dur00,fourP-1,fourP-2,fourP-3,fourP-4,fourP-5}. 
Unlike, however, three-qubit case their results seem to be contradictory to each other. In particular, the number of the SLOCC classes 
is different as Table I shows. Furthermore, we do not know any linear hierarchy in the four-qubit system.
Thus, our understanding on the four-qubit entanglement is still incomplete.

The purpose of this paper is to extend the analysis of Ref. \cite{elts12-1} to four-qubit system. For this purpose we choose nine SLOCC
classes of four-qubit system suggested in Ref. \cite{fourP-1}. This classification is achieved by making use of the Jordan block 
structure of some complex symmetric matrix. Nine classes and their representative states are 
\begin{eqnarray}
\label{slocc-ver}
G_{abcd}&=&\frac{a+d}{2}(|0000\rangle
+|1111\rangle)+\frac{a-d}{2}(|0011\rangle
+|1100\rangle)                                                  \nonumber  \\   
&&\hspace{.1cm}+\frac{b+c}{2}(|0101\rangle
+|1010\rangle)+\frac{b-c}{2}(|0110\rangle +|1001\rangle)                      \nonumber   \\  
L_{abc_2}&=&\frac{a+b}{2}(|0000\rangle
+|1111\rangle)+\frac{a-b}{2}(|0011\rangle +|1100\rangle)                             \nonumber   \\ 
&&\hspace{.5cm}+c(|0101\rangle +|1010\rangle)+|0110\rangle                            \nonumber   \\   
L_{a_2b_2}&=&a(|0000\rangle +|1111\rangle)+b(|0101\rangle+|1010\rangle)             
+|0110\rangle +|0011\rangle                                             \nonumber   \\
L_{ab_3}&=&a(|0000\rangle
+|1111\rangle)+\frac{a+b}{2}(|0101\rangle
+|1010\rangle)                                                                              \\
&&\hspace{.1cm}+\frac{a-b}{2}(|0110\rangle +|1001\rangle)                                   
+\frac{i}{\sqrt{2}}(|0001\rangle +|0010\rangle
+|0111\rangle
+|1011\rangle)                                                                               \nonumber    \\
L_{a_4}&=&a(|0000\rangle +|0101\rangle +|1010\rangle
+|1111\rangle)                                                                           
+(i|0001\rangle +|0110\rangle -i|1011\rangle)                            \nonumber    \\
L_{a_20_{3\oplus\bar{1}}}&=&a(|0000\rangle
+|1111\rangle)+(|0011\rangle +|0101\rangle +|0110\rangle)                                 \nonumber   \\
L_{0_{5\oplus\bar{3}}}&=&|0000\rangle +|0101\rangle +|1000\rangle
+|1110\rangle                                                                          \nonumber          \\
L_{0_{7\oplus\bar{1}}}&=&|0000\rangle +|1011\rangle +|1101\rangle
+|1110\rangle                                                                            \nonumber        \\
L_{0_{3\oplus\bar{1}}0_{3\oplus\bar{1}}}&=&|0000\rangle+|0111\rangle,                      \nonumber
\end{eqnarray}    
where $a$, $b$, $c$, and $d$ are complex parameters with nonnegative real part.

This paper is organized as follows. In sec. II we examine the four-qubit GHZ (GHZ$_4$) symmetry. Unlike the three-qubit case the 
whole set of GHZ$_4$-symmetric states is parametrized by three real parameters, {\it say} $x$, $y$, and $z$. In the parameter space 
all GHZ$_4$-symmetric states can be represented as points inside a tetrahedron. In sec. III we examine a question 
where $L_{abc_2}$, $L_{a_2b_2}$, 
$L_{a_20_{3\oplus\bar{1}}}$, $L_{0_{3\oplus\bar{1}}0_{3\oplus\bar{1}}}$, and $L_{a_4}$ GHZ$_4$-symmetric states reside 
in the tetrahedron, respectively. Using the results we derive the three linear hierarchies 
$L_{abc_2} \subset L_{a_4} \subset L_{a_2b_2} \subset G_{abcd}$, $L_{a_20_{3\oplus\bar{1}}} \subset G_{abcd}$, 
$L_{0_{3\oplus\bar{1}}0_{3\oplus\bar{1}}} \subset G_{abcd}$ which hold, at least, in the whole set of the GHZ$_4$-symmetric states.
Of course, these linear hierarchies are not complete because we have not analyzed other SLOCC classes ($L_{ab_3}$,
$L_{0_{5\oplus\bar{3}}}$, $L_{0_{7\oplus\bar{1}}}$) due to various difficulties. This difficulties are discussed in sec. IV. In the same 
section a brief conclusion is also given. In appendices A, B, C, D, and E we present a detailed calculation of sec. III, where Lagrange multiplier
technique is extensively used.

\section{GHZ$_4$ Symmetry}
It is straightforward to generalize the three-qubit GHZ symmetry to higher-qubit system. The direct generalization to four-qubit 
system can be written as a symmetry under (i) simultaneous flips (ii) qubit permutation (iii) qubit rotations about the $z$-axis
of the form
\begin{equation}
\label{four-1}
U (\phi_1, \phi_2, \phi_3) = e^{i \phi_1 \sigma_z} \otimes e^{i \phi_2 \sigma_z} \otimes e^{i \phi_3 \sigma_z}
                                \otimes e^{-i (\phi_1 + \phi_2 + \phi_3) \sigma_z}.
\end{equation}
One can show that the general form of the four-qubit states invariant under the transformations (i), (ii), and (iii) is 
\begin{eqnarray}
\label{4q-1}
& &\hspace{2.5cm} \rho_4^S = \beta \left[ \ket{0000}\bra{1111} + \ket{1111} \bra{0000} \right]                            \\    \nonumber
& &+ \mbox{diag} \left(\alpha_1, \alpha_2, \alpha_2, \alpha_3, \alpha_2, \alpha_3, \alpha_3, \alpha_2, \alpha_2, \alpha_3, \alpha_3, \alpha_2, 
                    \alpha_3, \alpha_2, \alpha_2, \alpha_1   \right)
\end{eqnarray}
where $\beta$, $\alpha_1$, $\alpha_2$ and $\alpha_3$ are real numbers satisfying $\alpha_1 + 4 \alpha_2 + 3 \alpha_3 = \frac{1}{2}$. 
Unlike the three-qubit case, $\rho_4^S$ is represented by three real parameters.

Now, we define the three real parameters $x$, $y$, $z$, as
\begin{eqnarray}
\label{change-1}
& &x = \beta                                                                 \\   \nonumber
& &y = \sqrt{\frac{8}{7}} \left(\alpha_1 - \frac{1}{16} \right)       \\   \nonumber
& &z = \sqrt{\frac{28}{3}} \left( \frac{\alpha_1}{7} + \alpha_2 - \frac{1}{14} \right).
\end{eqnarray}
Then, it is straightforward to show that the Hilbert-Schmidt metric of $\rho_4^S$ equals to the Euclidean metric, i.e.
\begin{equation}
\label{hilbert}
d^2 \left[ \rho_4^S (\alpha_1, \alpha_2, \alpha_3, \beta), \rho_4^S (\alpha_1', \alpha_2', \alpha_3', \beta') \right]
= (x - x')^2 + (y - y')^2 + (z - z')^2
\end{equation}
where $d^2 (A, B) = \frac{1}{2} \mbox{tr} (A-B)^{\dagger} (A-B)$.
The four-qubit GHZ states $\ket{\mbox{GHZ}}_{\pm} = (\ket{0000} + \ket{1111}) / \sqrt{2}$ correspond to 
$x = \pm 1/2$, $y = \sqrt{7 / 32}$, and $z=0$, respectively.

\begin{figure}[ht!]
\begin{center}

\includegraphics[height=5cm]{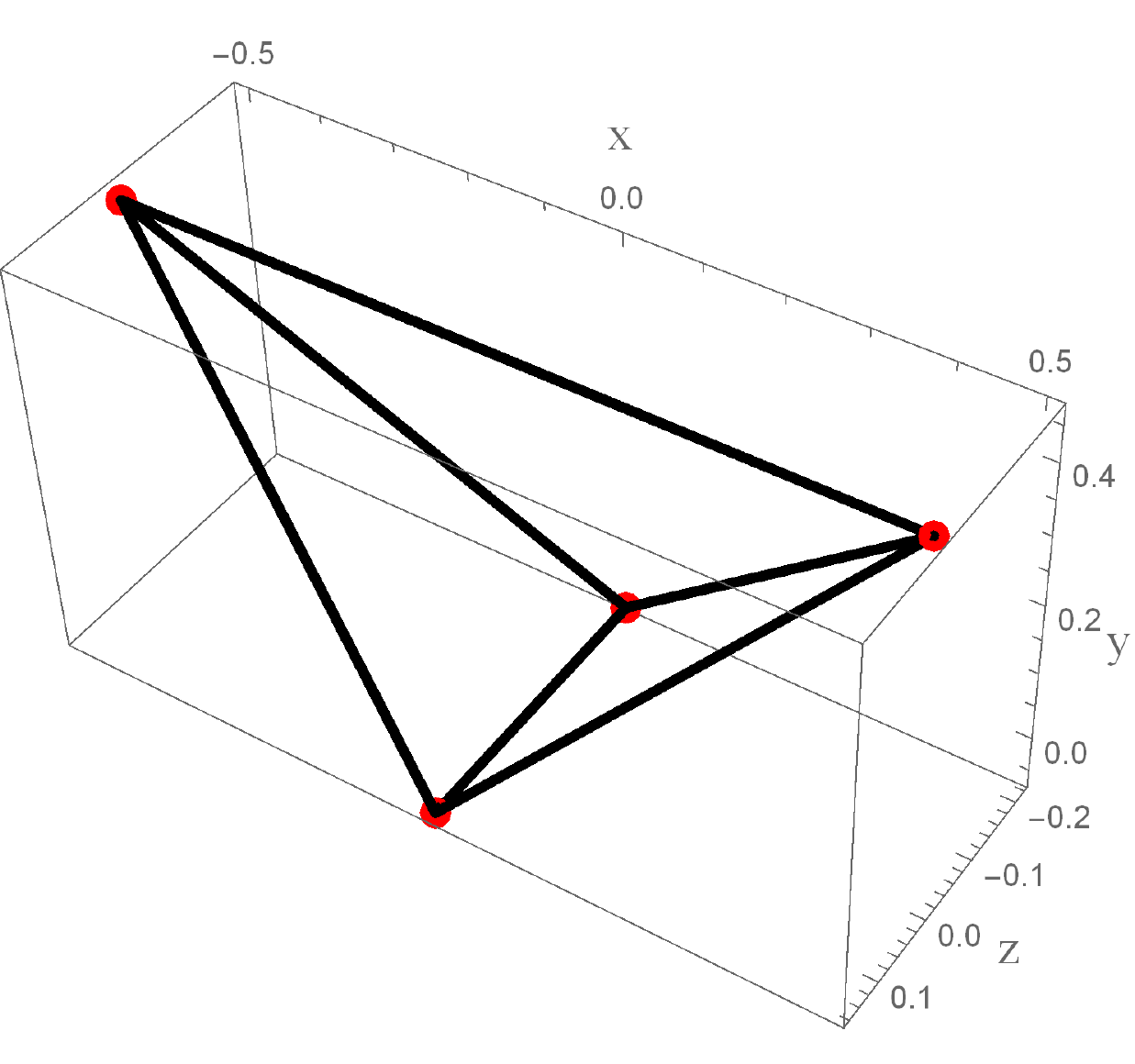}
\includegraphics[height=5cm]{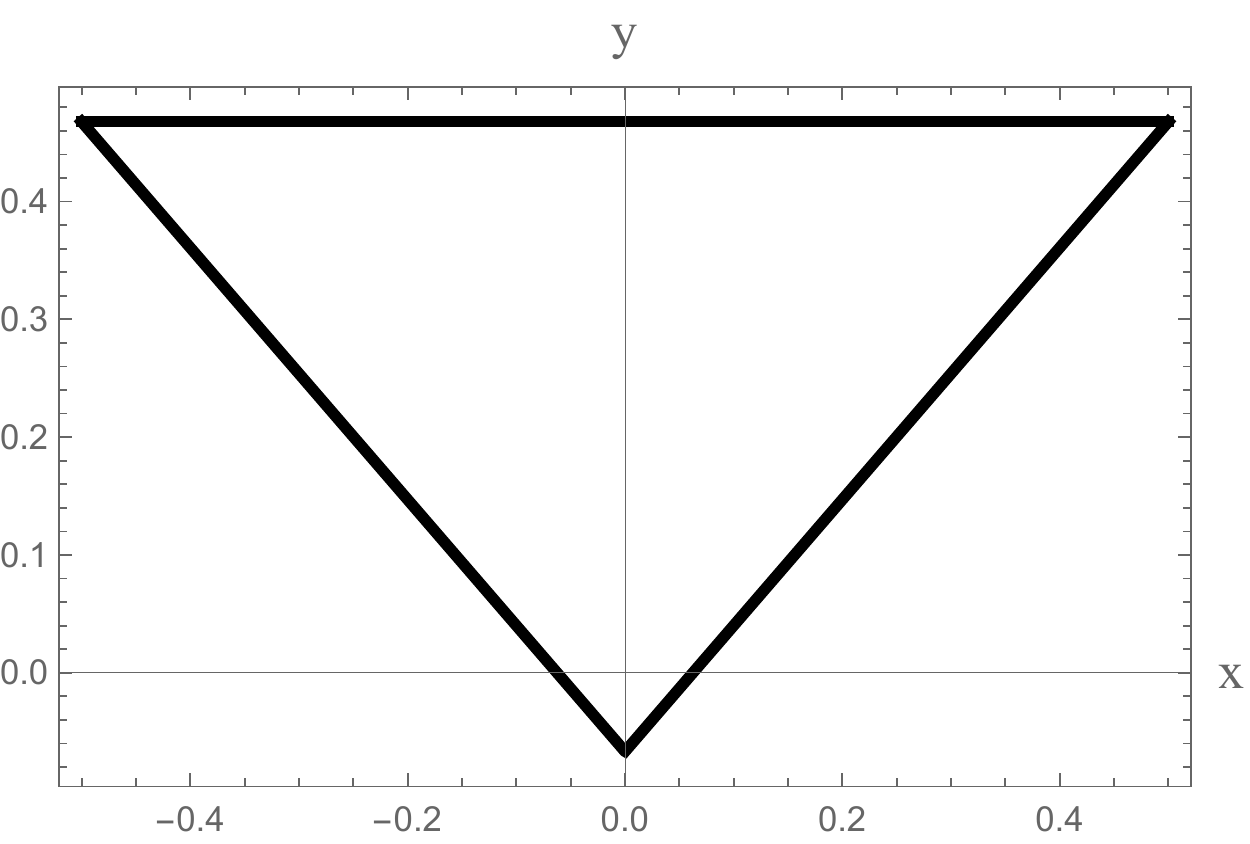}
\includegraphics[height=5cm]{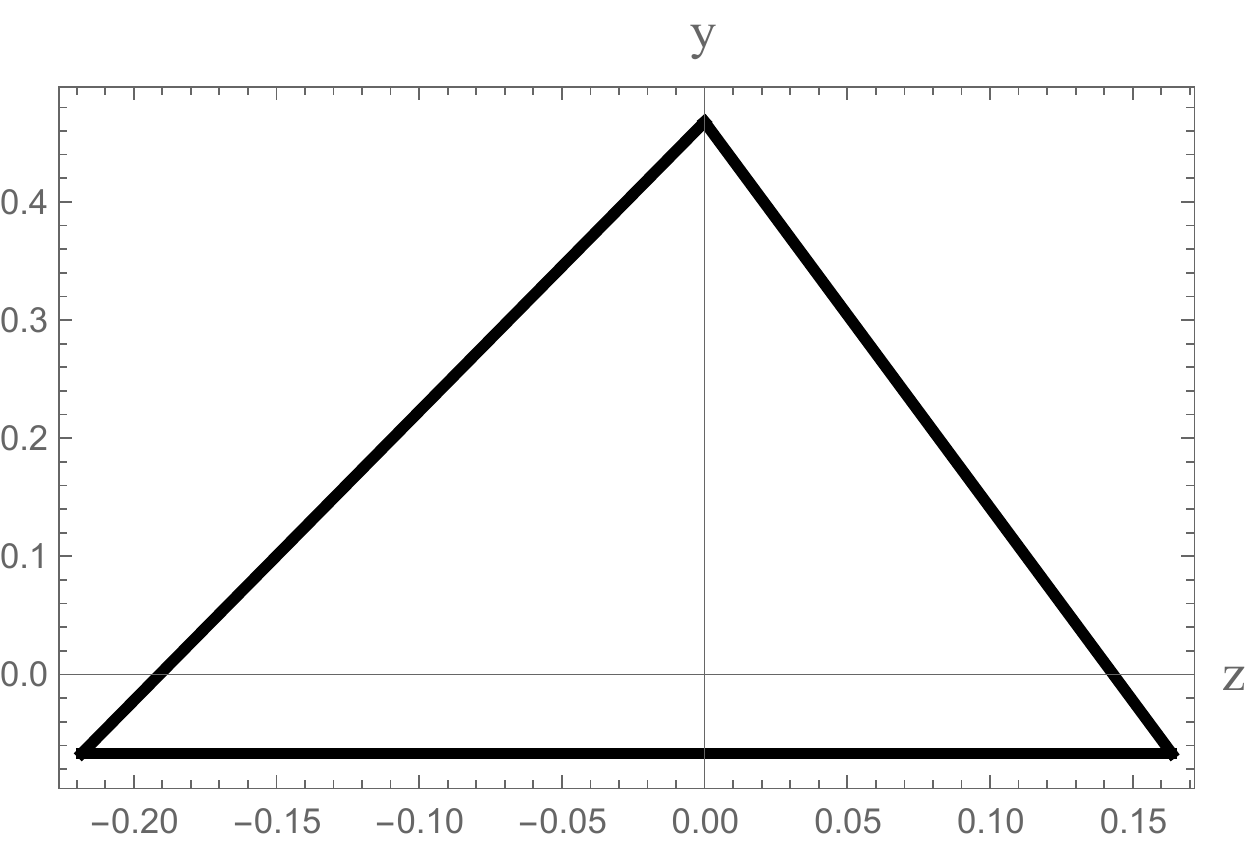}

\caption[fig2]{(Color online) (a) The restriction (\ref{restriction-1}) implies that all GHZ-symmetric states (\ref{4q-1}) should lie inside the tetrahedron.  
(b) Projection of the tetrahedron into ($x$, $y$)-plane. (c)  Projection of the tetrahedron into ($y$, $z$)-plane.}
\end{center}
\end{figure}

In order for $\rho_4^S$ to be a physical state we have the restrictions
\begin{equation}
\label{restriction-1}
0 \leq \alpha_2 \leq \frac{1}{8}                 \hspace{1.0cm}
0 \leq \alpha_3 \leq \frac{1}{6}                 \hspace{1.0cm}
0 \leq \alpha_1 \leq \frac{1}{2}                 \hspace{1.0cm}
0 \leq \alpha_1 \pm x \leq 1.
\end{equation}
Eq. (\ref{restriction-1}) implies that the physical state should lie inside the tetrahedron in
($x$, $y$, $z$)-spaces as Fig. 2(a) shows. The vertices of the tetrahedron and corresponding quantum states
are
\begin{eqnarray}
\label{vertices}
& &P_1 = \left(\frac{1}{2}, \sqrt{\frac{7}{32}}, 0 \right)   \hspace{.5cm}
\ket{\mbox{GHZ}_+} = \frac{1}{\sqrt{2}} \left(\ket{0000} + \ket{1111} \right)      \nonumber   \\
& &P_2 = \left(-\frac{1}{2}, \sqrt{\frac{7}{32}}, 0 \right)   \hspace{.5cm}
\ket{\mbox{GHZ}_-} = \frac{1}{\sqrt{2}} \left(\ket{0000} - \ket{1111} \right)      \nonumber   \\
& &P_3 = \left(0, -\frac{1}{8}\sqrt{\frac{2}{7}}, -\frac{1}{\sqrt{21}} \right)        \nonumber  \\
& &\pi_3 = \frac{1}{6} \bigg[ \ket{0011}\bra{0011} + \ket{0101}\bra{0101} + \ket{0110}\bra{0110} \\  \nonumber
& & \hspace{2.0cm}  + 
\ket{1001}\bra{1001} + \ket{1010}\bra{1010} + \ket{1100}\bra{1100}  \bigg]                \\   \nonumber
& &P_4 = \left(0, -\frac{1}{8}\sqrt{\frac{2}{7}}, \frac{\sqrt{21}}{28} \right)            \\   \nonumber
& &\pi_4 = \frac{1}{8} \bigg[ \ket{0001}\bra{0001} + \ket{0010}\bra{0010} + \ket{0100}\bra{0100} +
                     \ket{0111}\bra{0111}                                                  \\   \nonumber
& &  \hspace{2.0cm}
+ \ket{1000}\bra{1000} + \ket{1011}\bra{1011} + \ket{1101}\bra{1101} + \ket{1110}\bra{1110} \bigg].           
\end{eqnarray}
The origin in Fig. 2 (a) corresponds to the completely mixed state $I / 16$.
Eq. (\ref{restriction-1}) also implies that the projections of the tetrahedron into ($x$, $y$) and ($z$, $y$)
planes are Fig. 1(b) and Fig. 1(c), respectively. Thus, the physical states should reside in the triangles.
It is worthwhile noting that the sign of $x$ does not change the character of entanglement because 
$\rho_4^S (-x, y, z) = u \rho_4^S (x, y, z) u^{\dagger}$, where $u = i \sigma_x \otimes \sigma_y \otimes \sigma_y \otimes \sigma_y$.

Like a three-qubit GHZ symmetry there is a correspondence between four-qubit pure states and four-qubit GHZ$_4$-symmetric states as follows. Let 
$\ket{\psi}$ be a four-qubit pure state. Then, the corresponding GHZ$_4$-symmetric state $\rho_4^S (\psi)$ can be written as 
\begin{equation}
\label{correspondence-1}
\rho_4^S (\psi) = \int dU U \ket{\psi} \bra{\psi} U^{\dagger}
\end{equation}
where the integral is understood to cover the entire GHZ$_4$ symmetry group, i.e., unitaries $U (\phi_1, \phi_2, \phi_3)$ in Eq. (\ref{four-1})
and averaging over the discrete symmetries. For example, if $\ket{\psi} = \sum_{i,j,k,l=0}^1 \psi_{ijkl} \ket{ijkl}$, 
$\rho_4^S (\psi)$ becomes Eq. (\ref{4q-1}) with
\begin{eqnarray}
\label{correspondence-2}
& &x = \frac{1}{2} \left[ \psi_{0000} \psi_{1111}^* + \psi_{0000}^* \psi_{1111}\right]               \\   \nonumber
& &\alpha_1 = \frac{1}{2} \left[ |\psi_{0000}|^2 + |\psi_{1111}|^2 \right]                            \\   \nonumber
& &\alpha_2 = \frac{1}{8} \left[ |\psi_{0001}|^2 + |\psi_{0010}|^2 + |\psi_{0100}|^2 + |\psi_{1000}|^2 + |\psi_{1110}|^2 + |\psi_{1101}|^2 +
                                 |\psi_{1011}|^2 + |\psi_{0111}|^2 \right]                             \\   \nonumber
& &\alpha_3 = \frac{1}{6} \left[ |\psi_{0011}|^2 + |\psi_{0101}|^2 + |\psi_{0110}|^2 + |\psi_{1001}|^2 + |\psi_{1010}|^2 + |\psi_{1100}|^2 \right].
\end{eqnarray}
Note that $\bra{\psi} \psi\rangle = 1$ implies $2 \alpha_1 + 8 \alpha_2 + 6 \alpha_3 = 1$. 

\section{SLOCC classification of GHZ-symmetric states}
In this section we examine a question where $L_{abc_2}$, $L_{a_2b_2}$, 
$L_{a_20_{3\oplus\bar{1}}}$, $L_{0_{3\oplus\bar{1}}0_{3\oplus\bar{1}}}$, and $L_{a_4}$ GHZ$_4$-symmetric states reside 
in the tetrahedron of Fig. 2(a), respectively by applying the Lagrange multiplier method. The detailed calculation is presented in the five 
appendices. Similar issue was discussed in Ref. \cite{park14-1}. However, in this reference  the full GHZ$_4$ symmetry was not discussed 
because of calculation difficulties.

\subsection{$L_{abc_2}$}
The $L_{abc_2}$ SLOCC classification is represented as 
\begin{displaymath}
\label{labc2}
L_{abc_2} =\frac{a+b}{2}(|0000\rangle
+|1111\rangle)+\frac{a-b}{2}(|0011\rangle +|1100\rangle)                             
+c(|0101\rangle +|1010\rangle)+|0110\rangle  
\end{displaymath} 
where $a$ and $b$ are complex parameters with nonnegative real part.
In this section we want to explore a question where $L_{abc_2}$ GHZ$_4$-symmetric states reside in the tetrahedron.

The class $L_{abc_2}$ involves the fully separable state $\ket{0110}$ when $a=b=c=0$. In appendix A we use 
this state to show that when $y$ and $z$ are given, the $L_{abc_2}$ class of the GHZ$_4$-symmetric states resides
in $x \leq x_{max}$, where
\begin{equation}
\label{sep-1}
x_{max} = 2 \alpha_1 \frac{u}{1 + u^2}.
\end{equation}
In Eq. (\ref{sep-1}) $u$ is a quantity satisfying the quartic equation
\begin{equation}
\label{sep-2}
2 \left[ \alpha_1 (1 + u)^2 + 2 \alpha_2 (1 + u^2) \right]^2 - \alpha_1 (1 + u^2) (1 + u)^2 = 0.
\end{equation}

Eq. (\ref{sep-2}) can be solved numerically. The numerical result is presented in Fig. 3(a). 
The region where $L_{abc_2}$ GHZ$_4$-symmetric states reside can be described as follows. Consider a
rectangle $P_3 z_1 z_2 z_3$, where $P_3$ is given in Eq. (\ref{vertices}) and 
\begin{equation}
\label{sep-3}
z_1 = \left(\frac{1}{4}, \frac{3\sqrt{14}}{56}, -\frac{1}{2\sqrt{21}} \right)  \hspace{.5cm}
z_2 = \left(0, \sqrt{\frac{7}{32}}, 0 \right)                                  \hspace{.5cm}
z_3 = \left(-\frac{1}{4}, \frac{3\sqrt{14}}{56}, -\frac{1}{2\sqrt{21}} \right).
\end{equation}
Note that at these points $\alpha_2$ is zero. Now, bending this rectangle inward the tetrahedron, one can 
obtain the region where the GHZ$_4$-symmetric states of $L_{abc_2}$ reside. 

Note that even though we start with the fully separable state $\ket{0110}$, this region does not coincide
with the region where the PPT condition holds. Physically, this is because of the fact that $L_{abc_2}$ does
not involve only fully separable states. It contains entangled states depending on the parameters $a$, $b$, 
and $c$. Mathematically, this fact arises due to the fact that $x^{\Lambda}$ in Eq. (\ref{app-a-2}) has less symmetry due to $\Theta_2$ (see appendix A).

\begin{figure}[ht!]
\begin{center}

\includegraphics[height=5cm]{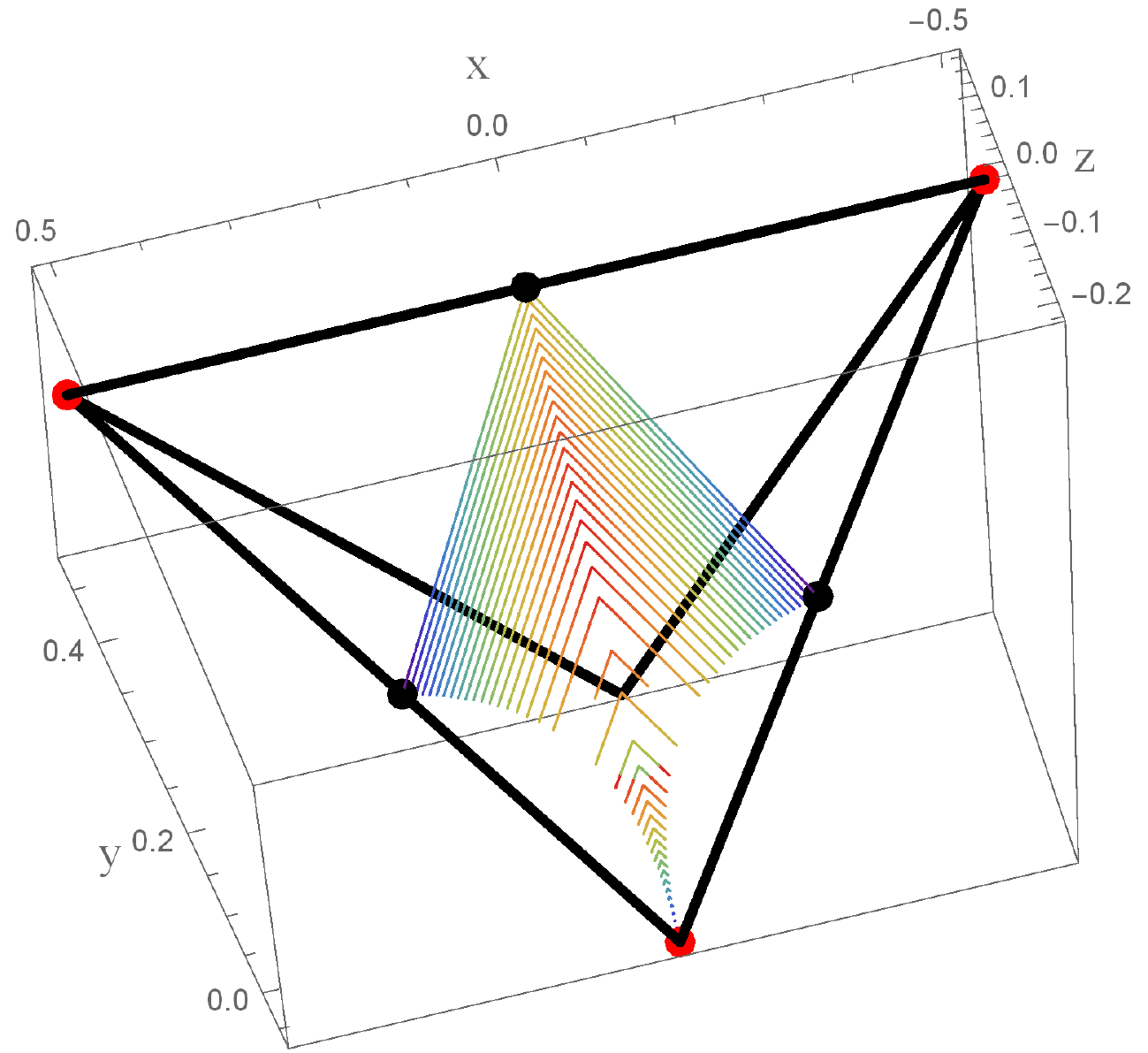}
\includegraphics[height=5cm]{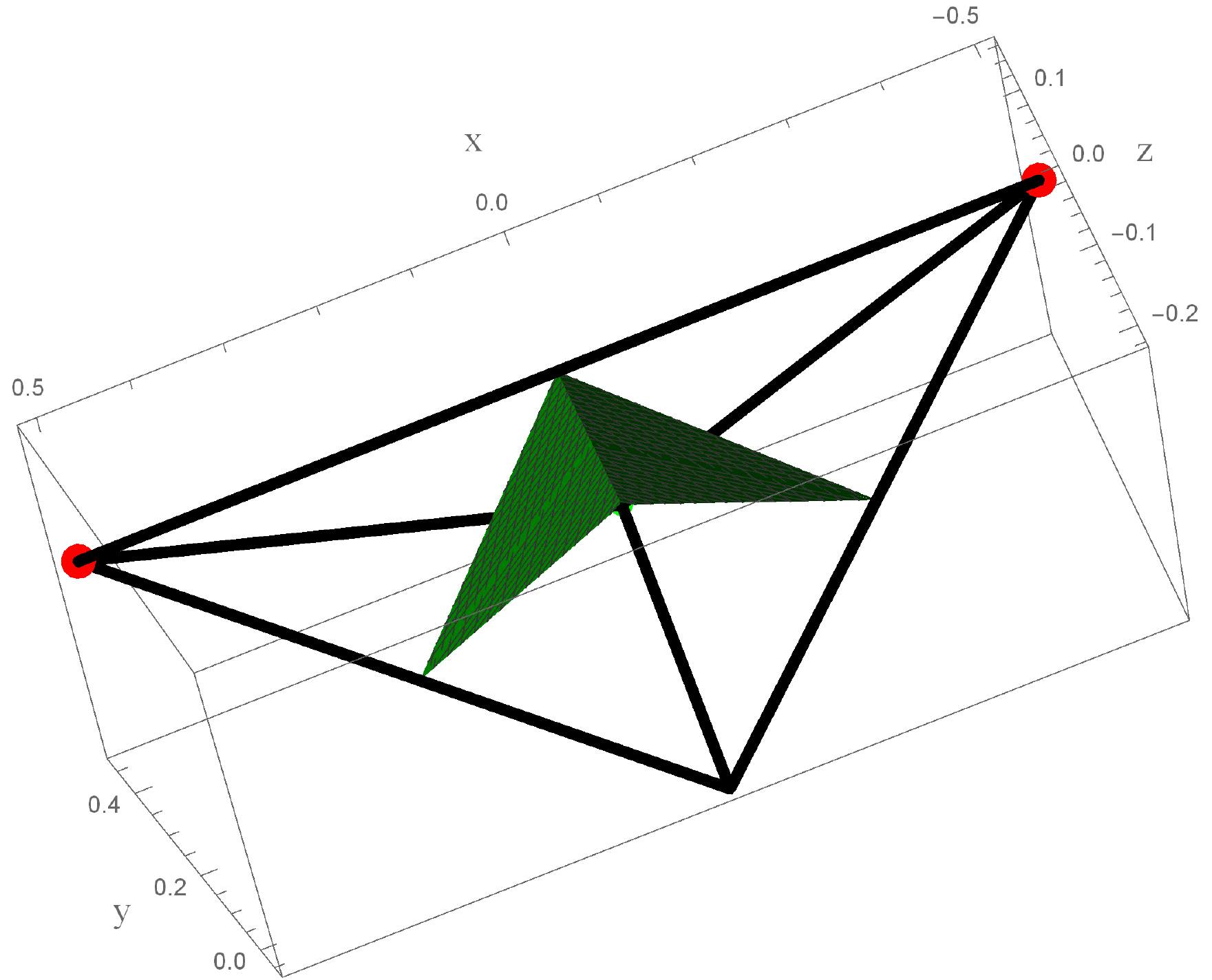}
\includegraphics[height=5cm]{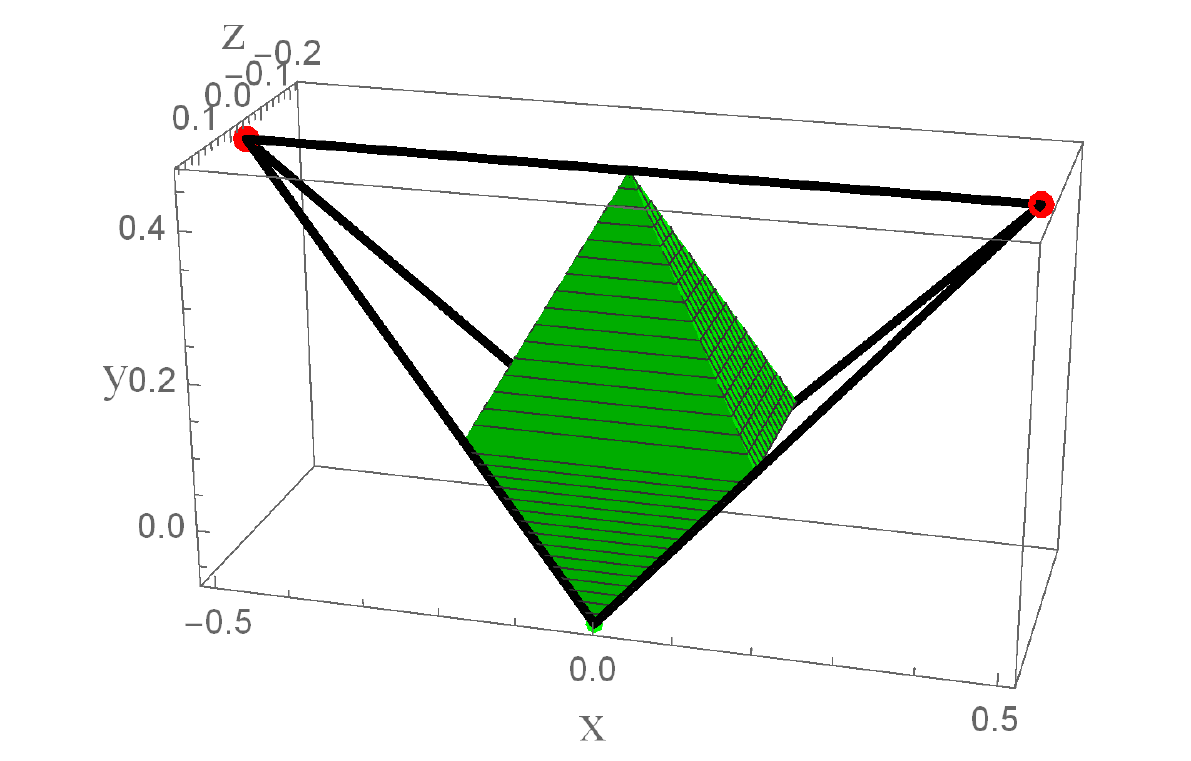}
\includegraphics[height=5cm]{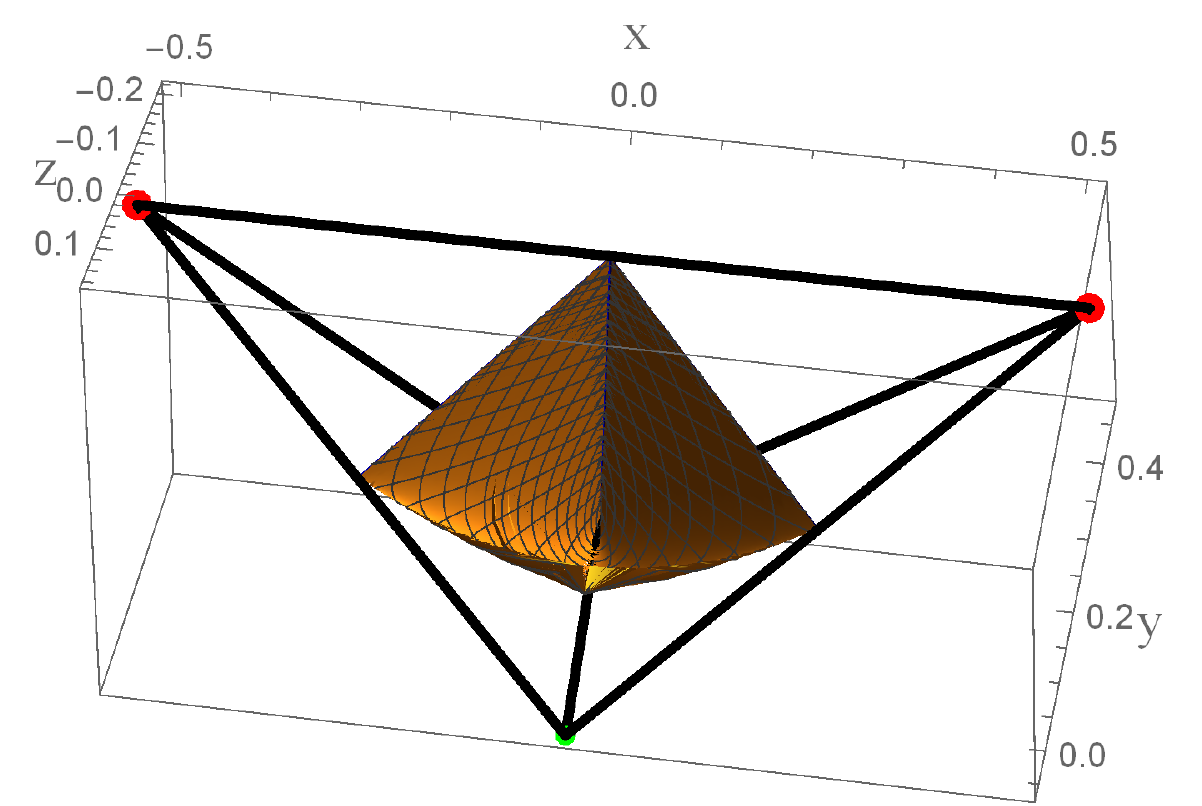}
\includegraphics[height=5cm]{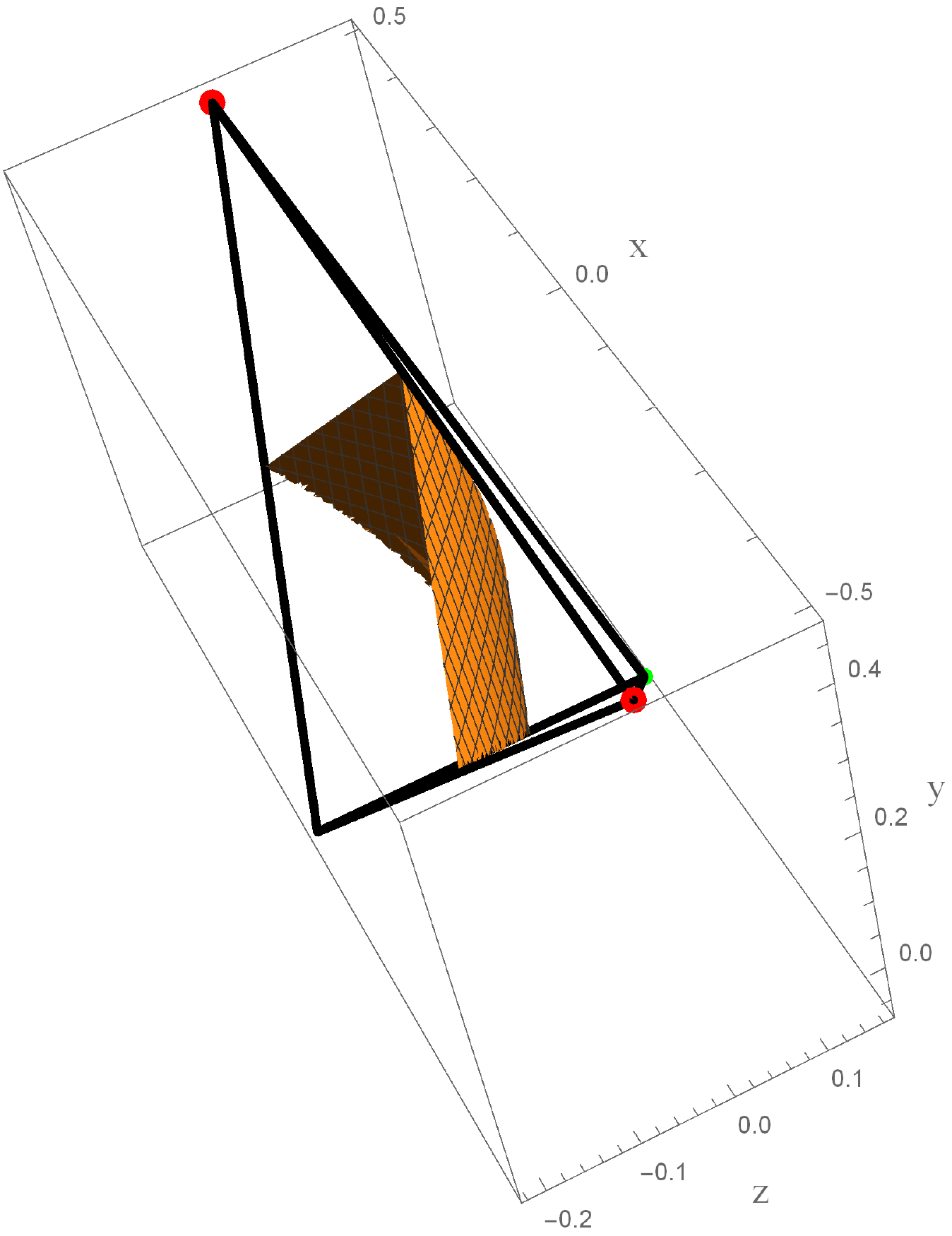}
\caption[fig3]{(Color online) The regions where the GHZ$_4$-symmetric states of (a) $L_{abc_2}$, (b) $L_{a_2b_2}$, 
(c) $L_{a_20_{3\oplus\bar{1}}}$, (d) $L_{0_{3\oplus\bar{1}}0_{3\oplus\bar{1}}}$, (e) $L_{a_4}$ classes reside 
in the tetrahedron of Fig. 2(a).   }
\end{center}
\end{figure}

\subsection{$L_{a_2b_2}$}
The $L_{a_2b_2}$ SLOCC classification is represented as 
\begin{displaymath}
\label{la2b2}
L_{a_2b_2}=a(|0000\rangle +|1111\rangle)+b(|0101\rangle+|1010\rangle)             
+|0110\rangle +|0011\rangle.
\end{displaymath}
In this section we want to explore a question where $L_{a_2b_2}$ GHZ$_4$-symmetric states reside in the tetrahedron.

The class $L_{a_2b_2}$ involves the partially separable state $\ket{0110} + \ket{0011}$ when $a=b=0$. 
In appendix B we use this state to show that when $y$ and $z$ are given, the $L_{a_2b_2}$ class of the GHZ$_4$-symmetric states resides in $x \leq x_{max}$, where $x_{max} = 3 \alpha_3$. This is represented in Fig. 3(b).
The remarkable fact is that the region represented by $x_{max} = 3 \alpha_3$ contains the region where 
$L_{abc_2}$ GHZ-symmetric states reside (Fig. 3(a)). This means the hierarchy $L_{abc_2} \subset L_{a_2b_2}$ holds, at least, in the GHZ$_4$-symmetric states.

\subsection{$L_{a_20_{3\oplus\bar{1}}}$}
The $L_{a_20_{3\oplus\bar{1}}}$ SLOCC classification is represented as 
\begin{displaymath}
\label{la2031}
L_{a_20_{3\oplus\bar{1}}}=a(|0000\rangle
+|1111\rangle)+(|0011\rangle +|0101\rangle +|0110\rangle).
\end{displaymath}
In this section we want to explore a question where $L_{a_20_{3\oplus\bar{1}}}$ GHZ$_4$-symmetric states reside in the
tetrahedron.

The class $L_{a_20_{3\oplus\bar{1}}}$ involves the partially separable state 
$\ket{0011} + \ket{0101} + \ket{0110}$ when $a=0$. 
In appendix C we use this state to show that when $y$ and $z$ are given, the $L_{a_20_{3\oplus\bar{1}}}$ class of the GHZ-symmetric states resides in $x \leq x_{max}$, where $x_{max} = 9 f \nu \mu_1 \mu_2^2$. The 
four parameters $f$, $\nu$, $\mu_1$, and $\mu_2$ satisfy
\begin{eqnarray}
\label{w-1}
& &3f(1+\nu^2) \left[3 + 3 \mu_1^2 \mu_2^4 + (\mu_1 + 2 \mu_2)^2 + \mu_2^2 (2 \mu_1 + \mu_2)^2 \right] = 1
                                                                                        \nonumber    \\
& &9 f (1 + \nu^2 \mu_1^2 \mu_2^4) = 2 \alpha_1               \\   \nonumber
& &3 f \left[3 \mu_1^2 \mu_2^4 + 3 \nu^2 + (\mu_1+2\mu_2)^2 +\nu^2 \mu_2^2(2\mu_1+\mu_2)^2 \right] = 8\alpha_2
                                                              \\    \nonumber
& &a \nu^4 + b \nu^2 + c = 0,
\end{eqnarray}
where
\begin{eqnarray}
\label{w-2}
& &a = (\mu_1+2\mu_2) (3 - 2\mu_1 \mu_2^3 - \mu_2^4)            \nonumber     \\
& &b = -(\mu_1^2 + \mu_1\mu_2 + \mu_2^2) (\mu_1 + 5 \mu_2 - 5 \mu_1 \mu_2^4 - \mu_2^5)     \\   \nonumber
& &c = \mu_1\mu_2^2 (2\mu_1 + \mu_2) (\mu_1 + 2 \mu_2 - 3 \mu_1 \mu_2^4).
\end{eqnarray}

From Eq. (\ref{w-1}) one can solve  $f$, $\nu$, $\mu_1$, and $\mu_2$ numerically. The region of the resulting
$x_{max}$ is not convex. Thus, we should choose the convex hull, which is represented by 
Fig. 3(c). The region where $L_{a_20_{3\oplus\bar{1}}}$ GHZ$_4$-symmetric states reside can be described as 
a polygon, whose vertices are $(0.185703, 0.13171, 0.102878)$, $(-0.185703, 0.13171, 0.102878)$, 
$(0, -(1/8) \sqrt{2 / 7} , - 1 / \sqrt{21})$, $(0.185703, 0.13171, -0.13717)$, 
$(-0.185703, 0.13171, -0.13717)$, and $(0, \sqrt{7/32}, 0)$. Comparing Fig. 3(c) with Fig. 3(a) and Fig. 3(b),
one can show that $L_{a_20_{3\oplus\bar{1}}}$ does not have any hierarchy relation with $L_{abc_2}$ and 
$L_{a_2b_2}$.

\subsection{$L_{0_{3\oplus\bar{1}}0_{3\oplus\bar{1}}}$}
The $L_{0_{3\oplus\bar{1}}0_{3\oplus\bar{1}}}$ SLOCC classification is represented as   
\begin{displaymath}
\label{l031031}
L_{0_{3\oplus\bar{1}}0_{3\oplus\bar{1}}}=|0000\rangle+|0111\rangle.
\end{displaymath}
In this section we want to explore a question where $L_{0_{3\oplus\bar{1}}0_{3\oplus\bar{1}}}$ GHZ$_4$-symmetric states
reside in the tetrahedron.

In appendix D we use this state to show that when $y$ and $z$ are given, the
$L_{0_{3\oplus\bar{1}}0_{3\oplus\bar{1}}}$ class of the GHZ$_4$-symmetric states resides in $x \leq x_{max}$,
where 
\begin{equation}
\label{ghz-1}
x_{max} = \sqrt{\frac{2 \alpha_1 (1 - 8 \alpha_2) - (1 - 16 \alpha_2 + 32 \alpha_2^2) + 6 \alpha_3 \sqrt{1 - 16 \alpha_2}}{2}}.
\end{equation}
However, the region generated by Eq. (\ref{ghz-1}) is not convex. Thus, we should choose its convex hull, 
which is depicted in Fig. 3(d). The region is composed of two plane triangles $z_2 P_3 r_1$ and $z_2 P_3 r_2$,
and a curved surface connecting these triangles, where 
$z_2$ and $P_3$ are given in Eq. (\ref{sep-3}) and Eq. (\ref{vertices}), respectively, and
\begin{equation}
\label{ghz-2}
r_1 = \left(\frac{1}{4}, \frac{3}{4 \sqrt{14}}, \frac{\sqrt{21}}{56} \right) \hspace{1.0cm}
r_2 = \left(-\frac{1}{4}, \frac{3}{4 \sqrt{14}}, \frac{\sqrt{21}}{56} \right).
\end{equation}
Comparing Fig. 3(d) with Fig. 3(a), Fig. 3(b) and Fig. 3(c),
one can show that $L_{0_{3\oplus\bar{1}}0_{3\oplus\bar{1}}}$ does not have any hierarchy relation with $L_{abc_2}$, 
$L_{a_2b_2}$, and $L_{a_20_{3\oplus\bar{1}}}$.

\subsection{$L_{a_4}$}
The $L_{a_4}$ SLOCC classification is represented as   
\begin{displaymath}
\label{a4-0}
L_{a_4} = a(|0000\rangle +|0101\rangle +|1010\rangle +|1111\rangle)                                                                           
+(i|0001\rangle +|0110\rangle -i|1011\rangle).
\end{displaymath}
When $a = 0$ this is reduced to $L_{a_4} = |0001\rangle +|0110\rangle -|1011\rangle$. The factor $i$ can be absorbed by redefining 
the second qubit as
$\ket {0} \rightarrow -i \ket{0}$. If we apply $\openone \otimes \openone \otimes \sigma_y \otimes \sigma_y$ and interchanging third and fourth
qubits,  $L_{a_4}$ reduces to 
\begin{equation}
\label{la4}
L_{a_4}=|0001\rangle+|0110\rangle+|1000\rangle.
\end{equation}
In this section we use this state to explore a question where $L_{a_4}$ GHZ$_4$-symmetric states
reside in the tetrahedron.

In appendix E we use this state to show that when $y$ and $z$ are given, the
$L_{a_4}$ class of the GHZ$_4$-symmetric states resides in $x \leq x_{max}$. where
\begin{equation}
\label{a4-1}
x_{max} = \frac{1}{2} \left[ (3 \alpha_3 - \alpha_1) + \sqrt{\frac{\alpha_1 + 3 \alpha_3 - 4 \alpha_2}{2}} \right].
\end{equation}
The region where the states of $L_{a_4}$-class is depicted in Fig. 3(e). Comparing this figure with other figures of Fig. 3 we can derive the 
linear hierarchy $L_{abc_2} \subset L_{a_4} \subset L_{a_2b_2}$, which holds, at least, in the GHZ$_4$-symmetric states.

\section{Conclusions}

In this paper we explore  GHZ$_4$ symmetry in four-qubit system. Unlike three-qubit GHZ symmetry the whole set of the 
GHZ$_4$-symmetric states is represented by three real parameters, {\it say} $x$, $y$, and $z$. In the parameter space 
all GHZ$_4$-symmetric states reside inside the tetrahedron of Fig. 2(a). 

Next, we explore a question where the given SLOCC class of the GHZ$_4$-symmetric states resides in the tetrahedron. Among nine SLOCC classes
we have examined five classes, i.e. $L_{abc_2}$, $L_{a_2b_2}$, $L_{a_20_{3\oplus\bar{1}}}$, $L_{0_{3\oplus\bar{1}}0_{3\oplus\bar{1}}}$, and $L_{a_4}$. Since the class $G_{abcd}$ involves the maximally entangled states,
it should be at the top in the linear hierarchy like GHZ class in three-qubit system.
Our analysis yields the following three different linear hierarchies 
$L_{abc_2} \subset L_{a_4} \subset L_{a_2b_2} \subset G_{abcd}$, $L_{a_20_{3\oplus\bar{1}}} \subset G_{abcd}$, and
$L_{0_{3\oplus\bar{1}}0_{3\oplus\bar{1}}} \subset G_{abcd}$, at least, in the whole set of the GHZ$_4$-symmetric states.
Of course, these linear hierarchies are incomplete because we have not analyzed the SLOCC classes $L_{ab_3}$,
$L_{0_{5\oplus\bar{3}}}$, and $L_{0_{7\oplus\bar{1}}}$ in the present paper. The reason why we have not analyzed these classes is 
mainly due to the following computational difficulties. The quantity $x^{\Lambda}$ defined in Eq. (\ref{app-a-2}) in 
$L_{ab_3}$, $L_{0_{5\oplus\bar{3}}}$, and $L_{0_{7\oplus\bar{1}}}$ classes has less symmetry than that in other SLOCC 
classes. Thus, computation of $x_{max}$ is highly complicated because of many free parameters in the Lagrange multiplier procedure.
Although we compute $x_{max}$ through numerical analysis, the resulting region in the tetrahedron becomes very complicated 
non-convex volume. Thus, it is highly difficult to derive the convex hull of this volume.

We hope to consider other numerical techniques, which enable us to treat the SLOCC classes 
$L_{ab_3}$, $L_{0_{5\oplus\bar{3}}}$, and $L_{0_{7\oplus\bar{1}}}$ in the future. If these techniques are available, it may 
lead the complete linear hierarchies in the four-qubit system.

{\bf Acknowledgement}:
This work was supported by the Kyungnam University Foundation Grant, 2016.

\newpage

\begin{appendix}{\centerline{\bf Appendix A: $L_{abc_2}$}}

\setcounter{equation}{0}
\renewcommand{\theequation}{A.\arabic{equation}}

In this appendix we prove Eq. (\ref{sep-1}) by applying the Lagrange multiplier method.
Let us define 
\begin{eqnarray}
\label{app-a-1}
\ket{\psi} = \left(  \begin{array}{cc}
                   A_1  &  B_1  \\
                   C_1  &  D_1
                     \end{array}        \right) \otimes
\left(  \begin{array}{cc}
                   A_2  &  B_2  \\
                   C_2  &  D_2
                     \end{array}        \right) \otimes
\left(  \begin{array}{cc}
                   A_3  &  B_3  \\
                   C_3  &  D_3
                     \end{array}        \right) \otimes
\left(  \begin{array}{cc}
                   A_4  &  B_4  \\
                   C_4  &  D_4
                     \end{array}        \right) \ket{0110}.
\end{eqnarray}
Then, it is straightforward derive the corresponding GHZ$_4$-symmetric state by making use of 
Eq. (\ref{correspondence-2}). In order to derive the maximum of $x$ when $y$ and $z$ are fixed,  
we define $x^{\Lambda}$ as 
\begin{equation}
\label{app-a-2}
x^{\Lambda} = x + \Lambda_0 \Theta_0 + \Lambda_1 \Theta_1 + \Lambda_2 \Theta_2,
\end{equation}
where
\begin{eqnarray}
\label{app-a-3}
& &x = (A_1 C_1) (B_2 D_2) (B_3 D_3) (A_4 C_4)       \nonumber    \\
& &\Theta_0 = (A_1^2 + C_1^2) (B_2^2 + D_2^2) (B_3^2 + D_3^2) (A_4^2 + C_4^2) - 1      \\   \nonumber
& &\Theta_1 = \frac{1}{2} \left[ (A_1 B_2 B_3 A_4)^2 + (C_1 D_2 D_3 C_4)^2 \right] - \alpha_1   \\ \nonumber  
& &\Theta_2 = \frac{1}{8} \left[ (A_1^2 D_2^2 + C_1^2 B_2^2) (B_3^2 A_4^2 + D_3^2 C_4^2) 
+ (A_1^2 B_2^2 + C_1^2 D_2^2) (B_3^2 C_4^2 + D_3^2 A_4^2)  \right] - \alpha_2.
\end{eqnarray}
Of course, the constraints $\Theta_j = 0 \hspace{.2cm} (j=1, 2, 3)$ arise from $\bra{\psi}\psi \rangle = 1$
and Eq. (\ref{correspondence-2}).

Note that $x$, $\Theta_0$, and $\Theta_1$ have $1 \leftrightarrow 4$ and $2 \leftrightarrow 3$ symmetries.
However, this symmetry does not hold in $\Theta_2$. Instead, it has $(1,2) \leftrightarrow (4,3)$ symmetry.
Thus, whole $x^{\Lambda}$ has   $(1,2) \leftrightarrow (4,3)$ symmetry. Using this symmetry $x_{max}$, maximum of $x$,
arises at $A_1^2 = A_4^2 \equiv a$, $C_1^2 = C_4^2 \equiv c$, $B_2^2 = B_3^2 \equiv b$, and 
$D_2^2 = D_4^2 \equiv d$. Then, $x$ and constraints become
\begin{eqnarray}
\label{app-a-4}
& &x = a b c d     \hspace{2.0cm}  \Theta_0 = (a+c)^2 (b+d)^2 - 1                 \\   \nonumber
& &\Theta_1 = \frac{1}{2} (a^2 b^2 + c^2 d^2) - \alpha_1            \hspace{1.0cm}
\Theta_2 = \frac{1}{4} (a b + c d) (a d + b c) - \alpha_2.
\end{eqnarray}

Now, one can derive the Lagrange equations 
$\frac{\partial x^{\Lambda}}{\partial z} = 0 \hspace{.2cm} (z= a, b, c, d)$ explicitly. 
However, we do not need these equations because the constraints $\Theta_j = 0$ fix $x$.
In order to show this let us define $\mu_1 = c / a$ and $\mu_2 = d / c$. And we define 
$v = \mu_1 + \mu_2$ and $u = \mu_1 \mu_2$. Then, $x = (a^2 b^2) u$ and the constraints become
\begin{equation}
\label{app-a-5}
(a^2 b^2) (1 + v + u)^2 = 1  \hspace{.5cm}
(a^2 b^2) (1 + u^2) = 2 \alpha_1    \hspace{.5cm}
(a^2 b^2) v (1 + u) = 4 \alpha_2.
\end{equation}
Eliminating $a^2 b^2$ and $v$ from Eq. (\ref{app-a-5}), one can derive Eq. (\ref{sep-2}). Also, combining 
Eq. (\ref{app-a-5}) and $x = (a^2 b^2) u$, one can derive Eq. (\ref{sep-1}).
\end{appendix}

\newpage

\begin{appendix}{\centerline{\bf Appendix B : $L_{a_2 b_2}$}}

\setcounter{equation}{0}
\renewcommand{\theequation}{B.\arabic{equation}}
In this appendix we prove that the $L_{a_2b_2}$ GHZ$_4$-symmetric states reside in the tetrahedron bounded by 
$x_{max} = 3 \alpha_3$ when $y$ and $z$ are fixed. Let us define 
\begin{eqnarray}
\label{app-b-1}
\ket{\psi} = \left(  \begin{array}{cc}
                   A_1  &  B_1  \\
                   C_1  &  D_1
                     \end{array}        \right) \otimes
\left(  \begin{array}{cc}
                   A_2  &  B_2  \\
                   C_2  &  D_2
                     \end{array}        \right) \otimes
\left(  \begin{array}{cc}
                   A_3  &  B_3  \\
                   C_3  &  D_3
                     \end{array}        \right) \otimes
\left(  \begin{array}{cc}
                   A_4  &  B_4  \\
                   C_4  &  D_4
                     \end{array}        \right) (\ket{0110} + \ket{0011}).
\end{eqnarray}
Then, the corresponding $x$ and $\alpha_j$ can be straightforwardly computed by making use of 
Eq. (\ref{correspondence-2}). Similar to appendix A we define
$x^{\Lambda}$ as Eq. (\ref{app-a-2}), where
\begin{eqnarray}
\label{app-b-2}
& &x = A_1C_1 B_3 D_3 (B_2 A_4 + A_2 B_4) (D_2 C_4 + C_2 D_4)     \nonumber    \\
& &\Theta_0 = (A_1^2 + C_1^2) (B_3^2 + D_3^2) \big[(B_2 A_4 + A_2 B_4)^2 + (B_2 C_4 + A_2 D_4)^2 \nonumber  \\
& &  \hspace{2.0cm}  + (D_2 A_4 + C_2 B_4)^2 + (D_2 C_4 + C_2 D_4)^2 \big] - 1      \\    \nonumber
& &\Theta_1 = \frac{1}{2} \left[ A_1^2 B_3^2 (B_2 A_4 + A_2 B_4)^2 + C_1^2 D_3^2 (D_2 C_4 + C_2 D_4)^2 \right]
                             - \alpha_1                                          \\    \nonumber
& &\Theta_2 = \frac{1}{8} \Bigg[ (A_1^2 B_3^2 + C_1^2 D_3^2) 
              \left\{ (B_2 C_4 + A_2 D_4)^2 + (D_2 A_4 + C_2 B_4)^2 \right\}       \\    \nonumber
& & \hspace{2.0cm}
+ (A_1^2 D_3^2 + C_1^2 B_3^2) \left\{ (B_2 A_4 + A_2 B_4)^2 + (D_2 C_4 + C_2 D_4)^2 \right\} \bigg] - \alpha_2.
\end{eqnarray}
Since $x^{\Lambda}$ has $2 \leftrightarrow 4$ symmetry, the maximum of $x$ occurs when
$A_2=A_4\equiv A$, $B_2=B_4\equiv B$, $C_2=C_4\equiv C$, and $D_2=D_4\equiv D$. Defining $\mu_1 = A B$, 
$\mu_2 = C D$, and $\mu_3 = A D + B C$, one can show that Eq. (\ref{app-b-2}) reduce to 
\begin{eqnarray}
\label{app-b-3}
& &x = 4 A_1 C_1 B_3 D_3 \mu_1 \mu_2              \nonumber    \\
& &\Theta_0 = 2 (A_1^2 + C_1^2) (B_3^2 + D_3^2) (2 \mu_1^2 + 2 \mu_2^2 + \mu_3^2) - 1     \\   \nonumber
& &\Theta_1 = 2 \left[ A_1^2 B_3^2 \mu_1^2 + C_1^2 D_3^2 \mu_2^2 \right] - \alpha_1       \\   \nonumber
& &\Theta_2 = \frac{1}{4} \left[(A_1^2 B_3^2 + C_1^2 D_3^2) \mu_3^2 + 2 (A_1^2 D_3^2 + C_1^2 B_3^2)
                                 (\mu_1^2 + \mu_2^2) \right] - \alpha_2.
\end{eqnarray}
Thus, we have seven Lagrange multiplier equations $\frac{\partial x^{\Lambda}}{\partial z} = 0 \hspace{.2cm}
(z = \mu_3, \mu_1, \mu_2, A_1, C_1, B_3, D_3)$ and three constraints $\Theta_j = 0 \hspace{.2cm} (j=0,1,2)$.
Among them $\frac{\partial x^{\Lambda}}{\partial \mu_3} = 0$ is solved by $\mu_3 = 0$. Using this solution, 
one can solve the remaining Lagrange multiplier equations. Finally, one can show that the Lagrange 
multiplier constants can be expressed in terms of the following ratios
\begin{equation}
\label{app-b-4}
g \equiv \frac{\mu_2}{\mu_1}  \hspace{.5cm}  r_1 \equiv \frac{C_1}{A_1}  \hspace{.5cm}
r_2 \equiv \frac{D_3}{B_3}.
\end{equation}
The explicit form of the Lagrange multiplier constants are 
\begin{eqnarray}
\label{app-b-5}
& &\Lambda_0 = \frac{r_2 (1 - g^2 r_1^2 r_2^2) (g^2 r_2^2 - r_1^2)}{2 g (1 + g^2) r_1 (1 - r_2^4) (1 - r_1^2 r_2^2)}
                                                                                      \nonumber   \\
& &\Lambda_1 = \frac{(1 - g^2) r_1 r_2}{g (1 - r_1^2 r_2^2)}                                \\   \nonumber
& &\Lambda_2 = \frac{4 r_2 (r_1^2 - g^2) (1 - g^2 r_1^2 r_2^2)}{g (1 + g^2) r_1 (1 - r_2^2) (1 - r_1^2 r_2^2)}.
\end{eqnarray}
Inserting Eq. (\ref{app-b-5}) into the remaining Lagrange multiplier equations, one can show that all
equation is solved by $g = r_1 r_2$. In terms of the ratios Eq. (\ref{app-b-3}) becomes
\begin{eqnarray}
\label{app-b-6}
& &x = 4 (A_1^2 B_3^2 \mu_1^2) g r_1 r_2                   \nonumber    \\
\Theta_0 = 0:  \hspace{.2cm}& & 4 (A_1^2 B_3^2 \mu_1^2) (1 + r_1^2) (1 + r_2^2) (1 + g^2) = 1     \\  \nonumber
\Theta_1 = 0: \hspace{.2cm}& & 2 (A_1^2 B_3^2 \mu_1^2) (1 + g^2 r_1^2 r_2^2) = \alpha_1    \\  \nonumber
\Theta_2 = 0: \hspace{.2cm}& & (A_1^2 B_3^2 \mu_1^2) (1 + g^2) (r_1^2 + r_2^2) = 2 \alpha_2.
\end{eqnarray}  
When $g = r_1 r_2$, one can show easily from Eq. (\ref{app-b-6}) that $x_{max}$ becomes
\begin{equation}
\label{app-b-7}
x_{max} = \frac{1}{2} (1 - 2 \alpha_1 - 8 \alpha_2) = 3 \alpha_3.
\end{equation}

Of course, there are many other solutions of the Lagrange multiplier equations. However, the resulting 
$x_{max}$ generated by other solutions are not physical. For example, Eq. (\ref{app-b-3}) with 
$g = 1/(r_1 r_2)$ also solve the Lagrange multiplier equations. In this case, $x_{max}$ becomes
$x_{max} = \alpha_1$. This means that all states in the tetrahedron are $L_{a_2b_2}$-class. Since,
however, $\ket{\mbox{GHZ}_{\pm}}$ are not  $L_{a_2b_2}$-class evidently, this solution is unphysical.

\end{appendix}
\newpage
\begin{appendix}{\centerline{\bf Appendix C : $L_{a_20_{3\oplus\bar{1}}}$}}

\setcounter{equation}{0}
\renewcommand{\theequation}{C.\arabic{equation}}
In this appendix we prove that the $L_{a_20_{3\oplus\bar{1}}}$ GHZ$_4$-symmetric states reside in the 
tetrahedron bounded by Eq. (\ref{w-1}).Let us define 
\begin{eqnarray}
\label{app-c-1}
& &\ket{\psi}                                   \\   \nonumber
& & = \left(  \begin{array}{cc}
                   A_1  &  B_1  \\
                   C_1  &  D_1
                     \end{array}        \right) \otimes
\left(  \begin{array}{cc}
                   A_2  &  B_2  \\
                   C_2  &  D_2
                     \end{array}        \right) \otimes
\left(  \begin{array}{cc}
                   A_3  &  B_3  \\
                   C_3  &  D_3
                     \end{array}        \right) \otimes
\left(  \begin{array}{cc}
                   A_4  &  B_4  \\
                   C_4  &  D_4
                     \end{array}        \right) (\ket{0011} + \ket{0101} + \ket{0110}).
\end{eqnarray}
Then, the corresponding $x$ and $\alpha_j$ can be computed by using Eq. (\ref{correspondence-2}).
Now, we define $x^{\Lambda}$ as Eq. (\ref{app-a-2}), where
\begin{eqnarray}
\label{app-c-2}
&&x=A_1C_1(A_2B_3B_4 + B_2A_3B_4 + B_2B_3A_4) (C_2D_3D_4 + D_2C_3D_4+D_2D_3C_4)        \nonumber  \\
& &\Theta_0 = (A_1^2+C_1^2) \Bigg[(A_2B_3B_4 + B_2A_3B_4 + B_2B_3A_4)^2 + (A_2B_3D_4 + B_2 A_3D_4+B_2B_3C_4)^2
                                                                                      \nonumber  \\
& &\hspace{2.0cm}+(A_2D_3B_4+B_2C_3B_4+B_2D_3A_4)^2 + (A_2D_3D_4+B_2C_3D_4+B_2D_3C_4)^2                 \nonumber  \\
&&\hspace{2.0cm}+(C_2B_3B_4+D_2A_3B_4+D_2B_3A_4)^2 + (C_2B_3D_4+D_2A_3D_4+D_2B_3C_4)^2      \nonumber \\
&&\hspace{1.0cm}+(C_2D_3B_4+D_2C_3B_4+D_2D_3A_4)^2 + (C_2D_3D_4+D_2C_3D_4+D_2D_3C_4)^2 \Bigg] - 1   
                                                                                             \\  \nonumber
&&\Theta_1=\frac{1}{2}\left[A_1^2(A_2B_3B_4+B_2A_3B_4+B_2B_3A_4)^2 + 
            C_1^2 (C_2D_3D_4+D_2C_3D_4+D_2D_3C_4)^2 \right] - \alpha_1                \\    \nonumber
&&\Theta_2=\frac{1}{8} \bigg[A_1^2  (A_2B_3D_4 + B_2 A_3D_4+B_2B_3C_4)^2 + 
                            A_1^2 (A_2D_3B_4+B_2C_3B_4+B_2D_3A_4)^2                   \\   \nonumber
&&\hspace{1.5cm}+A_1^2 (C_2B_3B_4+D_2A_3B_4+D_2B_3A_4)^2 + A_1^2 (C_2D_3D_4+D_2C_3D_4+D_2D_3C_4)^2    
                                                                                    \\   \nonumber
&&\hspace{1.5cm}+C_1^2 (A_2B_3B_4 + B_2A_3B_4 + B_2B_3A_4)^2 + C_1^2 (A_2D_3D_4+B_2C_3D_4+B_2D_3C_4)^2 
                                                                                     \\  \nonumber
&&\hspace{1.0cm}+C_1^2 (C_2B_3D_4+D_2A_3D_4+D_2B_3C_4)^2 + C_1^2 (C_2D_3B_4+D_2C_3B_4+D_2D_3A_4)^2 \Bigg] - \alpha_2.
\end{eqnarray}
Since $x^{\Lambda}$ has $2 \leftrightarrow 3$, $2 \leftrightarrow 4$, and $3 \leftrightarrow 4$
symmetries, the maximum of $x$ should occur at $A_2=A_3=A_4 \equiv A$, $B_2=B_3=B_4 \equiv B$,
$C_2=C_3=C_4 \equiv C$, and $D_2=D_3=D_4 \equiv D$. Then, Eq. (\ref{app-c-2}) reduces to
\begin{eqnarray}
\label{app-c-3}
&&x = 9 A_1C_1AB^2CD^2                    \\   \nonumber
&&\Theta_0=3(A_1^2+C_1^2) \left[3A^2B^4+3C^2D^4+B^2(2AD+BC)^2+D^2(AD+2BC)^2\right] - 1    \\  \nonumber
&&\Theta_1 = \frac{9}{2} \left[A_1^2A^2B^4 + C_1^2C^2D^4 \right] - \alpha_1               \\  \nonumber
&&\Theta_2 = \frac{3}{8} \left[3A_1^2C^2D^4 + 3C_1^2A^2B^4 + A_1^2B^2 (2AD+BC)^2+C_1^2D^2(AD+2BC)^2\right]
                            - \alpha_2.
\end{eqnarray}

Thus, we have six Lagrange multiplier equations 
$\frac{\partial x^{\Lambda}}{\partial z}=0 \hspace{.2cm}(z = A_1,C_1,A,B,C,D)$ and three constraints
$\Theta_j=0 \hspace{.2cm} (j=0, 1, 2)$. The six Lagrange multiplier equations are not all independent.
Now, we define 
\begin{equation}
\label{app-c-4}
\mu_1 = \frac{C}{A}   \hspace{1.0cm} \mu_2 = \frac{D}{B} \hspace{1.0cm} \nu= \frac{C_1}{A_1}.
\end{equation}
Then, $x$ and $\Theta_j=0$ becomes
\begin{eqnarray}
\label{app-c-5}
&&x = 9f \nu\mu_1\mu_2^2                   \nonumber   \\
&&3f (1 + \nu^2) \left[3 + 3 \mu_1^2\mu_2^4+(\mu_1+2\mu_2)^2+\mu_2^2 (2\mu_1+\mu_2)^2 \right]=1   \\  \nonumber
&&9 f (1+\nu^2\mu_1^2\mu_2^4) = 2 \alpha_1                     \\  \nonumber
&&3 f \left[3\mu_1^2\mu_2^4 + 3\nu^2 + (\mu_1+2\mu_2)^2 + \nu^2 \mu_2^2 (2\mu_1+\mu_2)^2 \right] = 8 \alpha_2.
\end{eqnarray}
where $f=A_1^2A^2B^4$. Eliminating $\Lambda_1$, one can derive the following two equations 
from the Lagrange multiplier:
\begin{equation}
\label{app-c-6}
2\Lambda_0 \beta_1 + \frac{\Lambda_2}{4} \beta_2 = 0
\hspace{1.0cm} 2\Lambda_0 \beta_3 + \frac{\Lambda_2}{4} \beta_4 = 0,
\end{equation}
where
\begin{eqnarray}
\label{app-c-7}
&&\beta_1 = (1 + \nu^2) \left[(\mu_1+2\mu_2) + \mu_2^2 (2\mu_1+\mu_2) \right]    \nonumber   \\
&&\beta_2 =  (\mu_1+2\mu_2) + \nu^2 \mu_2^2 (2\mu_1+\mu_2)            \\   \nonumber
&&\beta_3 = -\mu_1 \left\{(\mu_1+2\mu_2) + 2\mu_2^2 (2\mu_1+\mu_2) + 3\mu_1 \mu_2^4 \right\}
+ \nu^2 \left\{3 + 2\mu_2 (\mu_1+2\mu_2) + \mu_2^3 (2\mu_1+\mu_2) \right\}       \\   \nonumber
&&\beta_4=-\mu_1(3\mu_1\mu_2^4+\mu_1+2\mu_2) + \nu^2 (3+2\mu_1\mu_2^3+\mu_2^4).
\end{eqnarray}
Thus, the secular equation $\beta_1 \beta_4 - \beta_2 \beta_3 = 0$ becomes $a \nu^4 + b \nu^2 + c = 0$, where
the coefficients $a$, $b$, and $c$ are given in Eq. (\ref{w-2}).
\end{appendix}
\newpage
\begin{appendix}{\centerline{\bf Appendix D}}

\setcounter{equation}{0}
\renewcommand{\theequation}{D.\arabic{equation}}
In this appendix we prove Eq. (\ref{ghz-1}) by applying the Lagrange multiplier method. Let us define
\begin{equation}
\label{app-d-1}
\ket{\psi}                                  
 = \left(  \begin{array}{cc}
                   A_1  &  B_1  \\
                   C_1  &  D_1
                     \end{array}        \right) \otimes
\left(  \begin{array}{cc}
                   A_2  &  B_2  \\
                   C_2  &  D_2
                     \end{array}        \right) \otimes
\left(  \begin{array}{cc}
                   A_3  &  B_3  \\
                   C_3  &  D_3
                     \end{array}        \right) \otimes
\left(  \begin{array}{cc}
                   A_4  &  B_4  \\
                   C_4  &  D_4
                     \end{array}        \right) (\ket{0000} + \ket{0111}).
\end{equation}
Then, it is straightforward derive the corresponding GHZ$_4$-symmetric state by making use of 
Eq. (\ref{correspondence-2}). In order to apply the Lagrange multiplier method we define $x^{\Lambda}$ 
as Eq. (\ref{app-a-2}) with
\begin{eqnarray}
\label{app-d-2}
& &x = A_1C_1 (A_2A_3A_4+B_2B_3B_4)(C_2C_3C_4+D_2D_3D_4)              \nonumber    \\
&&\Theta_0 = (A_1^2+C_1^2) \bigg[(A_2A_3A_4+B_2B_3B_4)^2 + (A_2A_3C_4+B_2B_3D_4)^2 + (A_2C_3A_4+B_2D_3B_4)^2
                                                                           \nonumber   \\
&&\hspace{3.0cm}+ (A_2C_3C_4+B_2D_3D_4)^2 + (C_2A_3A_4+D_2B_3B_4)^2 + (C_2A_3C_4+D_2B_3D_4)^2  \nonumber \\
&&\hspace{4.0cm}+ (C_2C_3A_4+D_2D_3B_4)^2
+ (C_2C_3C_4+D_2D_3D_4)^2 \bigg] - 1                                                   \\   \nonumber
&&\Theta_1 = \frac{1}{2} \left[A_1^2(A_2A_3A_4+B_2B_3B_4)^2 + C_1^2(C_2C_3C_4+D_2D_3D_4)^2\right]-\alpha_1
                                                                                        \\   \nonumber
&&\Theta_2 = \frac{1}{8} \bigg[A_1^2(A_2A_3C_4+B_2B_3D_4)^2 + A_1^2(A_2C_3A_4+B_2D_3B_4)^2
+A_1^2(C_2A_3A_4+D_2B_3B_4)^2                                                            \\  \nonumber
&&\hspace{1.5cm} + A_1^2 (C_2C_3C_4+D_2D_3D_4)^2 + C_1^2(A_2A_3A_4+B_2B_3B_4)^2
+C_1^2(A_2C_3C_4+B_2D_3D_4)^2                                                             \\  \nonumber
&&\hspace{3.0cm} + C_1^2(C_2A_3C_4+D_2B_3D_4)^2 + C_1^2(C_2C_3A_4+D_2D_3B_4)^2 \bigg]-\alpha_2.
\end{eqnarray}
Since $x^{\Lambda}$ has $2 \leftrightarrow 3$, $2 \leftrightarrow 4$, and $3 \leftrightarrow 4$ symmetries,
the maximum of $x$ should occur at $A_2=A_3=A_4 \equiv A$, $B_2=B_3=B_4 \equiv B$, $C_2=C_3=C_4 \equiv C$, 
and $D_2=D_3=D_4 \equiv D$. Then, Eq. (\ref{app-d-2}) becomes
\begin{eqnarray}
\label{app-d-3}
&&x = A_1 C_1 (A^3 + B^3) (C^3 + D^3)                      \nonumber   \\
&&\Theta_0 = (A_1^2+C_1^2) \bigg[(A^3+B^3)^2 + 3 (A^2C+B^2D)^2        \nonumber  \\
&& \hspace{4.0cm} + 3(AC^2+BD^2)^2 + (C^3+D^3)^2 \bigg] - 1             \\   \nonumber
&&\Theta_1 = \frac{1}{2} \left[A_1^2 (A^3 + B^3)^2 + C_1^2 (C^3 + D^3)^2 \right] - \alpha_1    \\   \nonumber
&&\Theta_2 = \frac{1}{8} \bigg[3 A_1^2 (A^2 C + B^2 D)^2 + A_1^2 (C^3+D^3)^2              \\   \nonumber
&&\hspace{3.0cm}+ 3 C_1^2 (A C^2 + B D^2)^2
            + C_1^2 (A^3 + B^3)^2 \bigg] - \alpha_2.
\end{eqnarray}
Thus, we have six Lagrange multiplier equations 
$\frac{\partial x^{\Lambda}}{\partial z} = 0 \hspace{.2cm} (z = A_1, C_1, A, B, C, D)$.
Eliminating the Lagrange multiplier constants $\Lambda_j$, one can derive two relations
$A^2 C + B^2 D=0$ and $A C^2 + B D^2 = 0$.

Since both equations gives same $x_{max}$, we consider only $A^2 C + B^2 D=0$ in this appendix. This is solved
by $C = -\mu^2 D$ and $B = \mu A$. Defining $\nu = C_1 / A_1$ and $\rho=D/A$, one can re-express the 
four independent Lagrange multiplier equations in terms of $\mu$, $\nu$, and $\rho$:
\begin{eqnarray}
\label{app-d-4}
&&\nu\rho^3 (1 - \mu^3) + 2 \Lambda_0 \left[1 + 3\rho^4\mu^2+\rho^6(1-\mu^3)^2\right]+\Lambda_1 
+\frac{\Lambda_2}{4} \rho^6 (1-\mu^3)^2 = 0                   \\   \nonumber
&&\rho^3(1-\mu^3)+2\Lambda_0\nu  \left[1 + 3\rho^4\mu^2+\rho^6(1-\mu^3)^2\right]+\Lambda_1\nu\rho^6(1-\mu^3)^2
+ \frac{\Lambda_2}{4}\nu(1+ 3\rho^4\mu^2) = 0      \\   \nonumber
&&\nu\rho^3 (1 - \mu^3) + 2 \Lambda_0(1+\nu^2)(1+\rho^4\mu^5)+\Lambda_1 
+\frac{\Lambda_2}{4} \nu^2 (1+\rho^4 \mu^5) = 0         \\   \nonumber
&&\nu\mu+2 \Lambda_0(1+\nu^2) \left[-2\rho+\rho^3\mu(1-\mu^3)\right]+ \Lambda_1\nu^2\rho^3\mu(1-\mu^3)
+\frac{\Lambda_2}{4} \left[\rho^3\mu(1-\mu^3)-2\nu^2\rho\right]=0.
\end{eqnarray}
Three of Eq. (\ref{app-d-4}) can be used to derive the Lagrange multiplier constants. Since their 
explicit forms are lengthy, we do not present in this appendix. Eliminating all Lagrange multiplier 
constants from Eq. (\ref{app-d-4}), one can derive a relation
\begin{equation}
\label{app-d-5}
\rho^6 (1 - \mu^2)^2 = \nu^2 (1 + 3 \mu^2 \rho^4).
\end{equation}
In terms of $\mu$, $\nu$, and $\rho$, $x$ and $\Theta_j=0$ reduce to 
\begin{eqnarray}
\label{app-d-6}
&&x = f (1+\mu^3)^2\nu\rho^3(1-\mu^3)                \nonumber   \\
&&f(1+\mu^3)^2 \frac{(1+\nu^2)^2}{\nu^2}\rho^6 (1-\mu^3)^2=1          \\  \nonumber
&&f(1+\mu^3)^2 \left[1+\nu^2\rho^6 (1-\mu^3)^2\right] = 2\alpha_1      \\   \nonumber
&&f(1+\mu^3)^2\rho^6 (1-\mu^3)^2 = 4 \alpha_2,
\end{eqnarray}
where $f=A_1^2 A^6$. Combining Eq. (\ref{app-d-5}) and Eq. (\ref{app-d-6}), one can derive Eq. (\ref{ghz-1})
straightforwardly.

\end{appendix}

\newpage
\begin{appendix}{\centerline{\bf Appendix E}}

\setcounter{equation}{0}
\renewcommand{\theequation}{E.\arabic{equation}}
In this appendix we prove Eq. (\ref{a4-1}) by applying the Lagrange multiplier method. Let us define
\begin{equation}
\label{app-e-1}
\ket{\psi}                                  
 = \left(  \begin{array}{cc}
                   A_1  &  B_1  \\
                   C_1  &  D_1
                     \end{array}        \right) \otimes
\left(  \begin{array}{cc}
                   A_2  &  B_2  \\
                   C_2  &  D_2
                     \end{array}        \right) \otimes
\left(  \begin{array}{cc}
                   A_3  &  B_3  \\
                   C_3  &  D_3
                     \end{array}        \right) \otimes
\left(  \begin{array}{cc}
                   A_4  &  B_4  \\
                   C_4  &  D_4
                     \end{array}        \right) (\ket{0001} + \ket{0111} + \ket{1000}).
\end{equation}
Then, it is straightforward derive the corresponding GHZ$_4$-symmetric state by making use of 
Eq. (\ref{correspondence-2}). In order to apply the Lagrange multiplier method we define $x^{\Lambda}$ 
as Eq. (\ref{app-a-2}) with
\begin{eqnarray}
\label{app-e-2}
& &x = (A_1A_2A_3B_4 + A_1B_2B_3A_4 + B_1A_2A_3A_4)(C_1C_2C_3D_4 + C_1D_2D_3C_4 +D_1C_2C_3C_4)
                                                                                  \nonumber    \\
&&\Theta_0 = \bigg[(A_1A_2A_3B_4 + A_1B_2B_3A_4 + B_1A_2A_3A_4)^2 + (C_1A_2A_3B_4 + C_1B_2B_3A_4 + D_1A_2A_3A_4)^2                                     \nonumber    \\
&&\hspace{1.1cm}+ (A_1A_2A_3D_4 + A_1B_2B_3C_4 + B_1A_2A_3C_4)^2 + (C_1A_2A_3D_4 + C_1B_2B_3C_4 + D_1A_2A_3C_4)^2                                      \nonumber    \\
&&\hspace{1.1cm}+ (A_1A_2C_3B_4 + A_1B_2D_3A_4 + B_1A_2C_3A_4)^2 + (C_1A_2C_3B_4 + C_1B_2D_3A_4 + D_1A_2C_3A_4)^2                                      \nonumber    \\
&&\hspace{1.1cm}+ (A_1A_2C_3D_4 + A_1B_2D_3C_4 + B_1A_2C_3C_4)^2 + (C_1A_2C_3D_4 + C_1B_2D_3C_4 + D_1A_2C_3C_4)^2                                         \nonumber    \\
&&\hspace{1.1cm}+ (A_1C_2A_3B_4 + A_1D_2B_3A_4 + B_1C_2A_3A_4)^2 + (C_1C_2A_3B_4 + C_1D_2B_3A_4 + D_1C_2A_3A_4)^2                                         \nonumber    \\
&&\hspace{1.1cm}+ (A_1C_2A_3D_4 + A_1D_2B_3C_4 + B_1C_2A_3C_4)^2 + (C_1C_2A_3D_4 + C_1D_2B_3C_4 + D_1C_2A_3C_4)^2                                         \nonumber    \\
&&\hspace{1.1cm}+ (A_1C_2C_3B_4 + A_1D_2D_3A_4 + B_1C_2C_3A_4)^2 + (C_1C_2C_3B_4 + C_1D_2D_3A_4 + D_1C_2C_3A_4)^2                                          \nonumber    \\
&&\hspace{1.1cm}+ (A_1C_2C_3D_4 + A_1D_2D_3C_4 + B_1C_2C_3C_4)^2 + (C_1C_2C_3D_4 + C_1D_2D_3C_4 + D_1C_2C_3C_4)^2 \bigg]                                                 \nonumber   \\
&& \hspace{10cm} - 1                                                    \\   \nonumber
&&\Theta_1 = \frac{1}{2} \left[(A_1A_2A_3B_4 + A_1B_2B_3A_4 + B_1A_2A_3A_4)^2 + (C_1C_2C_3D_4 + C_1D_2D_3C_4 + D_1C_2C_3C_4)^2\right]                                                   \\   \nonumber
&& \hspace{10cm}-\alpha_1                                                \\   \nonumber
&&\Theta_2 = \frac{1}{8} \bigg[(A_1A_2A_3D_4 + A_1B_2B_3C_4 + B_1A_2A_3C_4)^2 + (A_1A_2C_3B_4 + A_1B_2D_3A_4 + B_1A_2C_3A_4)^2                                                          \\  \nonumber
&&\hspace{1.1cm}+ (A_1C_2A_3B_4 + A_1D_2B_3A_4 + B_1C_2A_3A_4)^2 + (A_1C_2C_3D_4 + A_1D_2D_3C_4 + B_1C_2C_3C_4)^2                                                             \\  \nonumber
&&\hspace{1.1cm}+ (C_1A_2A_3B_4 + C_1B_2B_3A_4 + D_1A_2A_3A_4)^2 + (C_1A_2C_3D_4 + C_1B_2D_3C_4 + D_1A_2C_3C_4)^2                                                              \\  \nonumber
&&\hspace{1.1cm}+ (C_1C_2A_3D_4 + C_1D_2B_3C_4 + D_1C_2A_3C_4)^2 + (C_1C_2C_3B_4 + C_1D_2D_3A_4 + D_1C_2C_3A_4)^2\bigg]                                                        \\  \nonumber
&& \hspace{10cm}-\alpha_2.
\end{eqnarray}
Since $x^{\Lambda}$ has $2 \leftrightarrow 3$ and $1 \leftrightarrow 4$ symmetries, the maximum of $x$
should occur at
\begin{eqnarray}
\label{app-e-3}
&&A_1=A_4\equiv A  \hspace{1.0cm} B_1=B_4\equiv B \hspace{1.0cm} C_1=C_4\equiv C \hspace{1.0cm}
D_1=D_4\equiv D                            \\    \nonumber
&&A_2=A_3\equiv \alpha  \hspace{1.0cm} B_2=B_3\equiv \beta  \hspace{1.0cm} C_2=C_3\equiv \gamma
 \hspace{1.0cm} D_2=D_3\equiv \delta.
\end{eqnarray}
Further, we define 
\begin{equation}
\label{app-e-4}
\mu_1 \equiv \frac{B}{A}  \hspace{1.0cm} \mu_2 \equiv \frac{D}{C}  \hspace{1.0cm}
\nu_1 \equiv \frac{\beta}{\alpha}  \hspace{1.0cm} \nu_2 = \frac{\delta}{\gamma}.
\end{equation}
Then, $x$ and $\Theta_j$ become
\begin{eqnarray}
\label{app-e-5}
&&x = (AC\alpha\gamma)^2 (2 \mu_1 + \nu_1^2) (2 \mu_2 + \nu_2^2)            \nonumber  \\
&&\Theta_0 = \left[ A^4 \gamma^4 (2 \mu_1 + \nu_2^2)^2 + C^4 \alpha^4 (2 \mu_2 + \nu_1^2)^2 + 
4 A^2C^2\alpha^2\gamma^2 (\mu_1 + \mu_2 + \nu_1 \nu_2)^2 \right] - 6 \alpha_3   \nonumber  \\
&&\Theta_1 = \left[A^4 \alpha^4 (2\mu_1 + \nu_1^2)^2 + C^4 \gamma^4 (2\mu_2 + \nu_2^2)^2 \right] - 2 \alpha_1
                                                                                 \\    \nonumber
&&\Theta_2 = \bigg[A^4\alpha^2 \gamma^2 (2\mu_1 + \nu_1\nu_2)^2 + C^4 \alpha^2 \gamma^2 (2\mu_2 + \nu_1\nu_2)^2
                                    \\   \nonumber
&& \hspace{3.0cm}
+ A^2C^2\alpha^4 (\mu_1 + \mu_2 + \nu_1^2)^2 + A^2C^2\gamma^4 (\mu_1 + \mu_2 + \nu_2^2)^2 \bigg] - 4\alpha_2.
\end{eqnarray}
In Eq. (\ref{app-e-5}) the constraint $\Theta_0$ is simplified by making use of $\Theta_1 = 0$, $\Theta_2 = 0$, 
and $2\alpha_1 + 8 \alpha_2 + 6\alpha_3 = 1$. If one defines again $a \equiv A^2$, $b \equiv B^2$, 
$c \equiv \alpha^2$, and $d = \gamma^2$, Eq. (\ref{app-e-5}) reduces to
\begin{eqnarray}
\label{app-e-6}
&&x = abcd  (2 \mu_1 + \nu_1^2) (2 \mu_2 + \nu_2^2)         \\   \nonumber
&&\Theta_0 = \left[a^2d^2  (2 \mu_1 + \nu_2^2)^2 + b^2c^2 (2 \mu_2 + \nu_1^2)^2 + 4 abcd  
(\mu_1 + \mu_2 + \nu_1 \nu_2)^2 \right] - 6 \alpha_3         \\   \nonumber
&&\Theta_1 = a^2c^2 (2\mu_1 + \nu_1^2)^2 + b^2d^2 (2\mu_2 + \nu_2^2)^2 - 2 \alpha_1     \\   \nonumber
&&\Theta_2 = \bigg[a^2cd (2\mu_1 + \nu_1\nu_2)^2 + b^2cd (2\mu_2 + \nu_1\nu_2)^2         \\   \nonumber
&&  \hspace{3.0cm}+ abc^2 (\mu_1 + \mu_2 + \nu_1^2)^2 + abd^2 (\mu_1 + \mu_2 + \nu_2^2)^2 \bigg] - 4\alpha_2. 
\end{eqnarray}

Now, one can derive eight Lagrange multiplier equations $\frac{\partial x^{\Lambda}}{\partial z} = 0
\hspace{.2cm} (z = a, b, c, d, \mu_1, \mu_2, \nu_1, \nu_2)$. From first four equations $(z = a, b, c, d)$, 
one can derive the Lagrange multiplier constants;
\begin{eqnarray}
\label{app-e-7}
&&2 \Lambda_0 = \frac{\Delta_1}{\Delta} b c d (2\mu_1 + \nu_1^2)(2 \mu_2 + \nu_2^2) (a b \omega_2 - c d \omega_1)                                                           \nonumber   \\
&&2 \Lambda_1 = -\frac{\Delta_2}{\Delta} b c d (2\mu_1 + \nu_1^2)(2 \mu_2 + \nu_2^2) (a b \omega_2 + c d \omega_1)                                                             \\    \nonumber
&&\Lambda_2 = \frac{\Delta_1 \Delta_2}{\Delta} b c d  (2\mu_1 + \nu_1^2)(2 \mu_2 + \nu_2^2)
\end{eqnarray}
where   
\begin{eqnarray}
\label{app-e-8}
&&\Delta_1 = a^2c^2  (2\mu_1 + \nu_1^2)^2 - b^2d^2 (2 \mu_2 + \nu_2^2)^2           \nonumber   \\
&&\Delta_2 = a^2d^2  (2\mu_1 + \nu_2^2)^2 - b^2c^2 (2 \mu_2 + \nu_1^2)^2           \\    \nonumber
&&\omega_1 = a^2 (2 \mu_1 + \nu_1\nu_2)^2 - b^2 (2\mu_2 + \nu_1\nu_2)^2           \\   \nonumber
&&\omega_2 = c^2 (\mu_1 + \mu_2 + \nu_1^2)^2 - d^2 (\mu_1 + \mu_2 + \nu_2^2)^2
\end{eqnarray}   
and 
\begin{eqnarray}
\label{app-e-9}
&&\Delta =  ac^2 (2\mu_1 + \nu_1^2)^2 (a b \omega_2 + c d \omega_1) \Delta_2
-  \left[ad^2 (2\mu_1 + \nu_2^2)^2 + 2bcd (\mu_1 + \mu_2 + \nu_1\nu_2)^2 \right] (a b \omega_2 - c d \omega_1)
\Delta_1                                                                       \nonumber   \\  
&& \hspace{2.0cm}
-\left[2 acd (2 \mu_1 + \nu_1\nu_2)^2 + bc^2  (\mu_1 + \mu_2 + \nu_1^2)^2 + bd^2 (\mu_1 + \mu_2 + \nu_2^2)^2
\right] \Delta_1 \Delta_2.
\end{eqnarray}  
Inserting Eq. (\ref{app-e-7}) into the remaining equations $(z = \mu_1, \mu_2, \nu_1, \nu_2)$,
one can derive four complicated equations. These equations are solved when $\nu_1=\nu_2 \equiv \nu$ and 
$\mu_1 = \mu_2 \equiv \mu$.  Then, $x$ and the constraints $\Theta_j = 0$ become
\begin{eqnarray}
\label{app-e-10}
&&x = (2\mu+\nu^2)^2 X Y          \hspace{1.0cm}
\frac{Z^4 + 4X Y Z^2 + X^2 Y^2}{Z^2} = f_0            \\   \nonumber
&&X^2 + Y^2 = f_1                  \hspace{1.0cm}
\frac{Z^2 + X Y}{Z} = f_2
\end{eqnarray}
where $X = a c$, $Y = b d$, $Z = a d$, and 
\begin{equation}
\label{app-e-11}
f_0 = \frac{6 \alpha_3}{(2 \mu + \nu^2)^2}               \hspace{1.0cm}
f_1 = \frac{2 \alpha_1}{(2 \mu + \nu^2)^2}               \hspace{1.0cm}
f_2 = \frac{4 \alpha_2}{(2 \mu + \nu^2)^2 (X+Y)}.
\end{equation}                              
From second and fourth equations of Eq. (\ref{app-e-10}), one can derive $Z^4 + (4 X Y - f_0)Z^2 + X^2 Y^2 = 0$
and $Z^2 = f_2 Z - X Y$. Combining these two equations, one can derive $f_2^2 - f_0 + 2 X Y = 0$, which finally
reduces to the quadratic equation
\begin{equation}
\label{app-e-12}
x^2 + (\alpha_1 - 3 \alpha_3) x + (4 \alpha_2^2 - 3 \alpha_1 \alpha_3) = 0.
\end{equation}
Thus, Eq. (\ref{a4-1}) is directly derived from Eq. (\ref{app-e-12}).
\end{appendix}

\end{document}